\documentclass[11pt]{llncs}
\usepackage{my-llncs}
\usepackage[normalem]{ulem}
\newcommand{\dlt}{\bgroup\markoverwith{\textcolor{red}{\rule[0.5ex]{1pt}{0.6pt}}}\ULon}

\newcommand{\Adv}{\ensuremath{\mathcal{A}}\xspace}

\newcommand{\algo}[1]{\ensuremath{\mathsf{#1}}\xspace}

\newcommand{\bits}{\ensuremath{\{0,1\}}\xspace}

\newcommand{\ct}{\ensuremath{\mathsf{ct}}\xspace}
\newcommand{\cind}{\ensuremath{\stackrel{\text{c}}{\approx}}\xspace}

\newcommand{\Dec}{\ensuremath{\mathsf{Dec}}\xspace}

\newcommand{\Enc}{\ensuremath{\mathsf{Enc}}\xspace}

\newcommand{\Eval}{\ensuremath{\mathsf{Eval}}\xspace}

\newcommand{\Gen}{\ensuremath{\mathsf{Gen}}\xspace}

\newcommand{\idind}{\ensuremath{\stackrel{\text{i.d.}}{=\mathrel{\mkern-3mu}=}}\xspace}

\newcommand{\KSam}{\ensuremath{\mathsf{KSam}}\xspace}

\newcommand{\LE}{\ensuremath{\mathsf{LE}}\xspace}

\newcommand{\ls}{\ensuremath{\mathsf{ls}}\xspace}

\newcommand{\msk}{\ensuremath{\mathsf{msk}}\xspace}

\newcommand{\Naturals}{\ensuremath{\mathbb{N}}\xspace}
\newcommand{\negl}{\ensuremath{\mathsf{negl}}\xspace}

\newcommand{\NP}{\ensuremath{\mathsf{NP}}\xspace}
\newcommand{\coNP}{\ensuremath{\mathsf{coNP}}\xspace}

\newcommand{\Output}{\ensuremath{\mathsf{Out}}\xspace}

\newcommand{\PI}{\ensuremath{\mathsf{PI}}\xspace}
\newcommand{\pick}{\ensuremath{\xleftarrow{\$}}\xspace}
\newcommand{\pk}{\ensuremath{\mathsf{pk}}\xspace}
\newcommand{\PKE}{\ensuremath{\mathsf{PKE}}\xspace}
\newcommand{\poly}{\ensuremath{\mathsf{poly}}\xspace}

\newcommand{\RndExt}{\ensuremath{\mathsf{RndExt}}\xspace}
\newcommand{\Ring}{\ensuremath{\mathsf{R}}\xspace}

\newcommand{\RS}{\ensuremath{\mathsf{RS}}\xspace}

\newcommand{\Set}[1]{\ensuremath{\{#1\}}\xspace}

\newcommand{\SecPar}{\ensuremath{\mathsf{\lambda}}\xspace}
\newcommand{\secpar}{\ensuremath{\mathsf{\lambda}}\xspace}

\newcommand{\Sig}{\ensuremath{\mathsf{Sig}}\xspace}
\newcommand{\Sign}{\ensuremath{\mathsf{Sign}}\xspace}

\newcommand{\sind}{\ensuremath{\stackrel{\text{s}}{\approx}}\xspace}
\newcommand{\sk}{\ensuremath{\mathsf{sk}}\xspace}
\newcommand{\SK}{\ensuremath{\mathsf{SK}}\xspace}

\renewcommand{\tilde}{\widetilde}

\newcommand{\Valid}{\ensuremath{\mathsf{Valid}}\xspace}

\newcommand{\Verify}{\ensuremath{\mathsf{Verify}}\xspace}

\newcommand{\VK}{\ensuremath{\mathsf{VK}}\xspace}

\newcommand{\xor}{\ensuremath{\mathsf{\oplus}}\xspace}

\newcommand{\ZAP}{\ensuremath{\mathsf{ZAP}}\xspace}

\renewcommand{\epsilon}{\varepsilon}

\newcommand{\PRF}{\ensuremath{\mathsf{PRF}}\xspace}
\newcommand{\PSF}{\ensuremath{\mathsf{PSF}}\xspace}
\newcommand{\Samp}{\ensuremath{\mathsf{Sample}}\xspace}
\newcommand{\F}{\ensuremath{\mathsf{F}}\xspace}
\newcommand{\InvF}{\ensuremath{\mathsf{F^{-1}}}\xspace}

\usepackage[absolute]{textpos}
\begin{document}

\begin{textblock}{105,90}(12,3)
\noindent A preliminary version of this paper appears in the proceedings of \textit{the 25th International Conference on Practice and Theory of Public-Key Cryptography} (\textsc{PKC 2022}). This is the full version.
\end{textblock}

\title{~\\~\\A Note on the Post-Quantum Security of (Ring) Signatures}
\author{
Rohit Chatterjee\inst{1} 
\and 
Kai-Min Chung\inst{2} 
\and 
Xiao Liang\inst{1}\thanks{Part of this work was done while visiting Max Planck Institute.} 
\and 
Giulio Malavolta\inst{3}}

\institute{
Stony Brook University, Stony Brook, USA\\
\email{rochatterjee@cs.stonybrook.edu}\\
\email{xiao.crypto@gmail.com} \\[0.5em]
\and
Academia Sinica, Taipei, Taiwan\\
\email{kmchung@iis.sinica.edu.tw}\\[0.5em]
\and
Max Planck Institute for Security and Privacy, Bochum, Germany\\
\email{giulio.malavolta@hotmail.it}
}

\let\oldaddcontentsline\addcontentsline
\def\addcontentsline#1#2#3{}
\maketitle
\def\addcontentsline#1#2#3{\oldaddcontentsline{#1}{#2}{#3}}



\pagenumbering{roman}
\phantomsection
\addcontentsline{toc}{section}{Abstract}
\begin{abstract}
This work revisits the security of classical signatures and ring signatures in a quantum world. For (ordinary) signatures, we focus on the arguably preferable security notion of {\em blind-unforgeability} recently proposed by Alagic et al.\ (Eurocrypt'20). We present two {\em short} signature schemes achieving this notion: one is in the quantum random oracle model, assuming quantum hardness of SIS; and the other is in the plain model, assuming quantum hardness of LWE with super-polynomial modulus. Prior to this work, the only known blind-unforgeable schemes are Lamport's one-time signature and the Winternitz one-time signature, and both of them are in the quantum random oracle model. 

\hspace{1em} For ring signatures, the recent work by Chatterjee et al.\ (Crypto'21) proposes a definition trying to capture adversaries with quantum access to the signer. However, it is unclear if their definition, when restricted to the classical world, is as strong as the standard security notion for ring signatures. They also present a construction that only {\em partially} achieves (even) this seeming weak definition, in the sense that the adversary can only conduct superposition attacks over the messages, but not the rings.
We propose a new definition that does not suffer from the above issue. Our definition is an analog to the blind-unforgeability in the ring signature setting. Moreover, assuming the quantum hardness of LWE, we construct a compiler converting any blind-unforgeable (ordinary) signatures to a ring signature satisfying our definition. 


\end{abstract}

\tableofcontents\addcontentsline{toc}{section}{Table of Contents}
\clearpage






\pagenumbering{arabic}

\section{Introduction}
Recent advances in quantum computing have uncovered several new threats to the  existing body of cryptographic work. As demonstrated several times in the literature (e.g., \cite{STOC:Watrous06,AC:BDFLSZ11,FOCS:Zhandry12,DBLP:conf/eurocrypt/AgarwalBGKM21}), building quantum-secure primitives requires more than taking existing constructions and replacing the underlying assumptions with post-quantum ones. It usually requires new techniques and analysis. Moreover, for specific primitives, even giving a meaningful security notion against quantum adversaries is a non-trivial task (e.g., \cite{EC:BonZha13,C:BonZha13,DBLP:journals/qic/Zhandry15,EC:Unruh16,EC:AMRS20}). This work focuses on {\em post-quantum security} of digital signature schemes, namely, classical signatures schemes for which we want to protect against quantum adversaries.

\para{Post-Quantum Unforgeable Signatures.} To build post-quantum secure signature schemes, the first step is to have a notion of unforgeability that protects against adversaries with quantum power. Probably the most natural attempt is to take the standard existential unforgeability (EUF) game, but require unforgeability against all {\em quantum polynomial-time} (QPT) adversaries (instead of all {\em probabilistic polynomial-time} (PPT) adversaries).  We emphasize that the communication between the EUF challenger and the QPT adversary is still classical. Namely, the adversary is not allowed to query the challenger's circuit in a quantum manner. Herein, we refer to this notion as PQ-EUF. Usually, PQ-EUF can be achieved by existing constructions in the classical setting via replacing the underlying hardness assumptions with quantum-hard ones (e.g., hard problems on lattice or isogeny-based assumptions). 

\subpara{The (Quantum) Random Oracle Model.} In the classical setting, the random oracle model (ROM) \cite{CCS:BelRog93} has been accepted as a useful paradigm to obtain efficient signature schemes. When considering the above PQ-EUF notion in the ROM, two choices arise---one can either allow the adversary {\em classical} access to the RO (as in the classical setting)\footnote{ To avoid confusion, we henceforth denote this model as CROM (``C'' for ``classical'').}, or {\em quantum} access to the RO. The latter was first formalized as the {\em quantum random oracle model} (QROM) by Boneh et al.\ \cite{AC:BDFLSZ11}, who showed that new techniques are necessary to achieve unforgeability against QPT adversaries in this model. Then, a large body of literature has since investigated the PQ-EUF in QROM \cite{FOCS:AmbRosUnr14,AC:Unruh17,EC:KilLyuSch18,C:DFMS19,C:LiuZha19,C:DonFehMaj20,EPRINT:GHHM20}.

\subpara{One-More Unforgeability vs Bind Unforgeability.} Starting from \cite{FOCS:Zhandry12}, people realize that the definitional approach taken by the above PQ-EUF may not be sufficient to protect against quantum adversaries. The reason is that quantum adversaries may try to attack the concerned protocol/primitive by executing it {\em quantumly}, even if the protocol/primitive by design is only meant to be executed classically. As argued in existing literature (e.g., \cite{DBLP:conf/icits/DamgardFNS13,C:GagHulSch16}), such an attack could possibly occur in a situation where the computer executing the classical protocol is a quantum machine, and an adversary somehow manages to observe the communication before measurement. Other examples include adversaries managing to trick a classical device (e.g., a smart card reader) into showing full or partial quantum behavior by, for example, cooling it down and shielding it from any external electromagnetic or thermal interference. Moreover, this concern may also arise in the security reduction (even) w.r.t.\  classical security games but against QPT adversaries. For example, some constructions may allow the adversary to obtain an {\em indistinguishability obfuscation} of, say, a PRF; the QPT adversary can then implement it as a quantum circuit to conduct superposition attacks. Recently, this issue has received an increasing amount of attention \cite{EC:BonZha13,C:BonZha13,DBLP:journals/qic/Zhandry15,EC:Unruh16,C:GagHulSch16,C:SonYun17,AC:HosYas18,AC:HosIwa19,C:CzaHulSch19,EC:AMRS20,DBLP:journals/iacr/CarstensETU20,DBLP:journals/iacr/ChevalierEV20,alagic2021impossibility,hosoyamada2021tight,DBLP:conf/crypto/HosoyamadaS21,cryptoeprint:2021:421}. 

To address the aforementioned security threats to digital signatures, it is reasonable to give the QPT adversary $\Adv$ {\em quantum access} to the signing oracle in the EUF game. This raises an immediate question---How should the game decide if $\Adv$'s final forgery is valid? Recall that in the classical setting (or the PQ-EUF above), the game records all the signing queries made by $\Adv$; to decide if $\Adv$ wins, it needs to make sure that $\Adv$'s final forgery message-signature pair is different from the ones $\Adv$ learned from the signing oracle. However, this approach does not fit into the quantum setting, since it is unclear how to record $\Adv$'s {\em quantum} queries without irreversibly disturbing them.

Boneh and Zhandry \cite{C:BonZha13} proposed the notion of {\em one-more unforgeability}. This requires that the adversary cannot produce $\mathsf{sq} + 1$ valid message-signature pairs with only $\mathsf{sq}$ signing queries (an approach previously taken to define blind signatures \cite{AC:PoiSte96}). When restricted to the classical setting, this definition is equivalent to the standard unforgeability of ordinary signatures, by a simple application of the pigeonhole principle. \cite{C:BonZha13} shows how to convert any PQ-EUF signatures to one-more unforgeable ones using a {\em chameleon hash function} \cite{DBLP:conf/ndss/KrawczykR00}; it also proves that the PQ-EUF signature scheme by Gentry, Peikert, and Vaikuntanathan \cite{STOC:GenPeiVai08} (henceforth, GPV) is one-more unforgeable in the QROM, assuming the PRF in that construction is quantum secure (i.e., being a QPRF \cite{FOCS:Zhandry12}).

However, as argued in \cite{C:GarYueZha17,EC:AMRS20}, one-more unforgeability does not seem to capture all that we can expect from quantum unforgeability. For example, an adversary may produce a forgery for a message in a subset $A$ of the message space, while making queries to the signing oracle supported on a disjoint subset $B$. Also, an adversary may make multiple quantum signing queries, but then must consume, say, all of the answers in order to make a single valid forgery. This forgery might be for a message that is different from all the messages in all the superpositions of previous queries. This clearly violates what we intuitively expect for unforgeability, but the one-more unforgeability definition may never rule this out. 

To address these problems, Alagic at el.\ \cite{EC:AMRS20} propose {\em blind-unforgeability} (BU). Roughly, the blind-unforgeability game modifies the (quantum-accessible) signing oracle by asking it to always return ``$\bot$'' for messages in a ``blinded'' subset of the message space. The adversary's forgery is considered valid only if it lies in the blinded subset. In this way, the adversary is forced to forge a signature for a message she has not seen a signature before, consistent with our intuition for unforgeability. \cite{EC:AMRS20} shows that blind-unforgeability, when restricted to the classical setting, is also equivalent to PQ-EUF; Moreover, it does not suffer from the above problems for one-more unforgeability\footnote{Although \cite{EC:AMRS20} claimed that blind-unforgeability implies one-more unforgeability, their proof was flawed \cite{BU:vs:OneMore}. The relation between these two notions is still an open problem.}.

In terms of constructions, \cite{EC:AMRS20} show that Lamport's one-time signature \cite{lamport1979constructing} is BU in the QROM, assuming the OWF is modeled as a (quantum-accessible) random oracle. Later, \cite{majenz2021quantum} show that the Winternitz one-time signature \cite{C:Merkle89a} is BU in the QROM, assuming the underlying hash function is modeled as a (quantum-accessible) random oracle. To the best of our knowledge, these are the only schemes known to achieve BU. This gives rise to the following question:
\begin{quote}
{\bf Question 1:}
{\em Is it possible to build (multi-time) signature schemes achieving blind-unforgeability, either in the QROM or the plain model?}
\end{quote}
{\bf Post-Quantum Secure Ring Signatures.} In a {\em ring signature} scheme \cite{AC:RivShaTau01,TCC:BenKatMor06}, a user can sign a message with respect to a {\em ring} of public keys, with the knowledge of a signing key corresponding to any public key in the ring. It should satisfy two properties: 
\begin{enumerate}
\item
 {\em Anonymity} requires that no user can tell which user in the ring actually produced a given signature; 
 \item
  {\em Unforgeability} requires that no user outside the specified ring can produce valid signatures on behalf of this ring. 
\end{enumerate}
  In contrast to its notional predecessor, {\em group signatures} \cite{EC:ChaVan91}, no central coordination is required for producing and verfying ring signatures. Due to these features, ring signatures (and their variants) have found natural applications related to whistleblowing, authenticating leaked information, and more recently to cryptocurrencies \cite{torres2018post,EPRINT:Noether15}, and thus have received extensive attention (see, e.g., \cite{C:CGHKLMPS21} and related work therein). 


For ring signatures from {\em latticed-based} assumptions,
there exist several constructions in the CROM \cite{AFRICACRYPT:ABBFG13,EC:LLNW16,torres2018post,DBLP:conf/icics/BaumLO18,DBLP:journals/ijhpcn/WangZZ18,CCS:EZSLL19,AC:BeuKatPin20,DBLP:conf/crypto/LyubashevskyNS21}, but only two schemes are known in the plain model \cite{DBLP:journals/iacr/BrakerskiK10,C:CGHKLMPS21}. The authors of \cite{C:CGHKLMPS21} also initiate the study of quantum security for ring signatures. They propose a definition where the QPT adversary is allowed quantum access to the signing oracle in both the anonymity and unforgeability game, where the latter is a straightforward adaption of the aforementioned one-more unforgebility for ordinary signatures. As noted in their work, this approach suffers from two disadvantages: 
\begin{enumerate}
\item
Their unforgeability definition seems weak in the sense that, when restricted to the classical setting, it is unclear if their unforgeability is equivalent to the standard one (see \Cref{sec:tech-overview:PQ-ring-sig}). This is in contrast to ordinary signatures, for which one-more unforgeability is equivalent to the standard existential unforgeability;
\item 
Their construction only partially achieves (even) this seemingly weak definition. In more detail, their security proof only allows the adversary to conduct superposition attacks on the messages, but not on the rings. As remarked by the authors, this is not a definitional issue, but rather a limitation of their technique. Indeed, \cite{C:CGHKLMPS21} left it as an open question to have a construction protecting against superposition attacks on both the messages and the rings.
\end{enumerate} 

The outlined gap begs the following natural question:
\begin{quote}
{\bf Question 2:} {\em Can we have a proper unforgeability notion for ring signatures that does not suffer from the above disadvantage? If so, can we have a construction achieving such a notion?}
\end{quote}

\para{Our Results.} In this work, we resolve the aforementioned questions:
\begin{enumerate}
\item
We show that the GPV signature, which relies on the quantum hardness of SIS (QSIS), can be proven BU-secure in the QROM. Since our adversary has quantum access to the signing oracle, we also need to replace the PRF in the original GPV scheme with a QPRF, which is also known from QSIS. As will be discussed later in \Cref{sec:tech-overview:BU:QROM}, our security proof is almost identical to the proof in \cite{C:BonZha13} for the one-more unforgeability of GPV, except how the desired contradiction is derived in the last hybrid. Interestingly, our proof for BU turns out to be simpler than that in \cite{C:BonZha13} (for one-more unforgeability). We remark that the GPV scheme is {\em short} (i.e., the signature size only depends on the security parameter, but not the message size).

\item
We also construct a BU-secure signature {\em in the plain model}, assuming  quantum hardness of Learning with Errors (QLWE) with super-polynomial modulus. Our construction is inspired by the signature (and adaptive IBE) scheme by Boyen and Li \cite{AC:BoyLi16}. This signature scheme is also short.

\item \label{item:contribution:ring-sig-def}
We present a new definition of post-quantum security for ring signatures, by extending blind-unforgeability from \cite{EC:AMRS20}. We show that this definition, when restricted to the classical setting, is equivalent to the standard security requirements for ring signatures.

\item We build a ring signature satisfying the above definition. Our construction is a compiler that converts any BU (ordinary) signature to a ring signature achieving the definition in \Cref{item:contribution:ring-sig-def}, assuming QLWE. 
\end{enumerate}

\section{Technical Overview}

\subsection{BU Signatures in the QROM}\label{sec:tech-overview:BU:QROM}
We show that the GPV signature scheme from \cite{STOC:GenPeiVai08} is BU-secure in the QROM. 
The GPV signature scheme follows the hash-and-sign paradigm and relies crucially on the notion of {\em preimage sampleable functions} (PSFs). As the name indicates, these functions can be efficiently inverted given a secret inverting key in addition to being efficiently computable. Further, the joint distribution of image-preimage pairs is statistically close, no matter whether the image or the preimage is sampled first. PSFs also provide collision resistance, as well as {\em pre-image min-entropy}: given any image, the set of possible preimages has $\omega(\log \secpar)$ bits of min-entropy, meaning that a specific preimage can only be predicted with negligible chance. 

The GPV scheme uses a hash function $H$ modeled as a random oracle. It first hashes the message $m$ using $H$ to obtain a digest $h$. The signing key includes the PSF secret key, and the signature is a preimage of $h$ (the signing randomness is generated using a quantum secure PRF over the message). 
To verify a signature, one simply computes its image under the PSF and compares it with the digest. 

Notice that in the proof of (post-quantum) blind-unforgeability, the adversary has quantum access to both $H$ and the signing algorithm. To show blind-unforgeability, we will move to a hybrid experiment where the $H$ and the signing algorithm $\Sign$ are constructed differently, but their {\em joint distribution} is statistically close to that in the real execution. To do so, the hybrid will set the signature for a message $m$ to a random preimage from the domain of the PSF (note that this procedure is ``de-randomized'' using the aforementioned PRF). To answer a $H$-oracle query on $m$, the hybrid will first compute its signature (i.e., the PSF preimage corresponding to $m$), and then return the PSF evaluation on this signature (aka preimage) as the output of $H(m)$. Observe that, in this hybrid, the $(H, \Sign)$ oracles are constructed by first sampling preimages for the PSF, and then evaluating the PSF in the ``forward'' direction; in contrast, in the real game, the $(H, \Sign)$ oracles can be interpreted as sampling a image for PSF first, and then evaluating the PSF in the ``reverse'' direction using the inverting key. From the property of PSFs given above, these two approaches induce statistically-close joint distributions of $(H, \Sign)$ on each (classical) query. A lemma from \cite{C:BonZha13} then shows that these are also indistinguishable to adversaries making polynomially-many {\em quantum} queries. 

So far, our proof is identical to that of \cite{C:BonZha13}, where GPV is shown to be one-more unforgeable. This final part is where we differ. In the final hybrid, if the adversary produces a successful forgery for a message in the blind set, only two possibilities arise. Since the image of the signature under the PSF must equal the digest, the signature must either (i) provide a second preimage for $h$ to the one computed by the challenger, creating a collision for the PSF, or (ii) equal the one the challenger itself computes, compromising preimage min-entropy of the PSF. This latter claim requires special attention in \cite{C:BonZha13}. A reduction to the min-entropy condition is not immediate, since it is unclear if the earlier quantum queries of $\Adv$ already allow $\Adv$ information about the preimages for the $q+1$ forgeries it outputs. To handle this, \cite{C:BonZha13} prove a lemma (\cite[Lemma 2.6]{C:BonZha13}) showing $q$ quantum queries will not allow $\Adv$ to predict $q+1$ preimages, given the min-entropy condition. In contrast, this last argument is superfluous in our case, since the blind unforgeability game {\em automatically} prevents any information for queries in the blindset from reaching the adversary. We can therefore directly appeal to the min-entropy condition for case (ii) above. 

We present the formal construction and the corresponding proof in \Cref{sec:BU:sig:QROM}.

\subsection{BU Signatures in the Plain Model}

To construct a BU signature {\em in the plain model}, we make use of the signature template introduced in \cite{AC:BoyLi16}, which in turn relies on key-homomorphic techniques as used in \cite{ITCS:BraVai14}. We will refer to the \cite{ITCS:BraVai14} homomorphic evaluation procedure as $\algo{Eval}_{\textsc{bv}}$. The \cite{AC:BoyLi16} scheme uses the `left-right trapdoor' paradigm. Namely, the verification key contains a matrix $\vb{A}$ sampled with a `trapdoor' basis $\vb{T}_{\vb{A}}$, and $\vb{A}_0,\vb{C}_0,\vb{A}_1,\vb{C}_1$, which can be interpreted as BV encodings of 0 and 1 respectively, as well as similar encodings $\Set{\vb{B}_i}_{i \in [|k|]}$ of the bits of a key $k$ for a bit-PRF (the use of this PRF is the key innovation in \cite{AC:BoyLi16}). The corresponding signing key contains $\vb{T}_{\vb{A}}$.  To sign, one computes   BV encodings $\mathbf{C}_{M_1},\dots,\mathbf{C}_{M_t}$ of a $t$-bit message $M$, then computes $\vb{A}_{\textsc{prf,m}} = \algo{Eval}_{\textsc{bv}}(\Set{\vb{B}_i}_{i \in [|k|]},\Set{\vb{C}_j}_{j \in [t]},\algo{PRF})$. Two signing matrices $\mathbf{F}_{M,b} = [\mathbf{A}~|~\mathbf{A}_b - \mathbf{A}_\textsc{prf,m}]$ ($\forall b \in \Set{0,1}$) are then generated (crucially, the adversary cannot tell these apart because of the PRF). A signature is a {\em short non-zero} vector $\sigma \in \mathbb{Z}^{2m}$ satisfying $\mathbf{F}_{M,b}\cdot\sigma= 0$ for any one of the $\mathbf{F}_{M,b}$'s. As pointed out, $\vb{T}_{\vb{A}}$ allows the signer to produce a short vector for either $\mathbf{F}_{M,b}$.

To show unforgeability, one constructs a reduction that 
\begin{enumerate}
\item
replaces the left matrix with an SIS challenge (thus losing $\vb{T}_{\vb{A}}$), {\bf and}
\item
 replaces the other matrices used to generate the right half with their `puncturable' versions (e.g., $\mathbf{A}_b$ now becomes $\mathbf{AR}_b + \mathbf{G}$, where $\mathbf{R}_b$ is an uniform low-norm matrix and $\mathbf{G}$ is the gadget matrix), with the end result being that the matrix $\vb{A}_{\textsc{prf,m}}$ becomes $\vb{A}\vb{R'} + \vb{G}$ and $\mathbf{F}_{M,b}$ now looks like $[\mathbf{A}~|~\mathbf{AR}+(b-\algo{PRF}_k(M))\mathbf{G}]$ (with $\vb{R},\vb{R'}$ being suitable low-norm matrices). 
\end{enumerate} 
The crucial point is this: having sacrificed $\vb{T}_{\vb{A}}$, the reduction cannot sign like a normal signer. However it still retains a trapdoor for the gadget matrix $\mathbf{G}$, and for {\em exactly one} of the $\mathbf{F}_{M,b}$, a term in $G$ survives in the right half. This suffices to obtain a `right trapdoor', and in turn, valid signatures for any $M$. On the other hand, a forging adversary lacks the PRF key and so it cannot tell apart $\mathbf{F}_{M,0}$ from $\mathbf{F}_{M,1}$. Thus the forgery must correspond to $\mathbf{F}_{M,\algo{PRF}_k(M)}$ with probability around $1/2$, and the reduction can use this solution to obtain a short solution for the challenge $\mathbf{A}$. 

However, the blind-unforgeability setting differs in several meaningful ways. Here, we no longer expect a forgery for any possible message, so the additional machinery to have two signing matrices for every message becomes superfluous. Indeed, for us the challenge is to disallow signing queries in the blindset (even if they are made as part of a query superposition) and to prevent forgeries in the blindset. Accordingly, we interpret the function of the PRF in a different manner. We simply have the bit-PRF act as the characteristic function for the blindset. Then we can extend the approach above to the blind-unforgeability setting very easily: we use a single signing matrix $\mathbf{F}_\textsc{m} = [\mathbf{A}~|~\mathbf{A'} - \mathbf{A}_\textsc{prf,m}]$ (where $\mathbf{A'}$ `encodes 1').  In the reduction, after making changes just as before, we obtain that $\mathbf{F}_\textsc{m} = \big[\mathbf{A}~|~\mathbf{AR} - \big(1-\algo{PRF}_k(M)\big)\mathbf{G}\big]$. For messages where the PRF is not 1, we can answer signing queries using the trapdoor for $\vb{G}$; For messages where it is 1, we cannot, and further we can use a forgery for such a message to break the underlying SIS challenge. In effect, the reduction enforces the requisite blindset behavior naturally.

A caveat is that the bit-PRF based approach may not correctly model a blindset, which is a random $\epsilon$-weight set of messages. Indeed, we require a slight modification of a normal bit-PRF to allow us the necessary latitude in approximating sets of any weight $\epsilon \in [0,1]$. Moreover, due to the adversary's quantum access to the signing oracle, this PRF must be quantum-access secure; and to allow the BV homomorphic evaluation, the PRF must have $\mathsf{NC}^1$ implementation. Fortunately, such a {\em biased} bit-PRF can be built by slightly modifying the PRF from \cite{EC:BanPeiRos12}, assuming QLWE with super-polynomial modulus.




\subsection{Post-Quantum Secure Ring Signatures}
\label{sec:tech-overview:PQ-ring-sig}
{\bf Defining Post-Quantum Security.} To reflect the {\em quantum power} of an QPT adversary $\Adv$, one needs to give $\Adv$ quantum access to the signing oracle in the security game. While this is rather straightforward for anonymity, the challenge here is to find a proper notion for unforgeability (thus, here we only focus on the latter). Let us first recall the {\em classical} unforgeability game for a ring signature. In this game, $\Adv$ learns a ring $\mathcal{R}$ from the challenger, and then can make two types of queries: 
\begin{itemize}
\item
 by a {\em corruption query} $(\algo{corrupt}, i)$, $\Adv$ can corrupt a member in $\mathcal{R}$ to learn its secret key; 
 \item
 by a {\em signing query} $(\algo{sign}, i, \Ring^*, m)$, $\Adv$ can create a ring $\Ring^*$, specify a member $i$ that is contained in both $\mathcal{R}$ and $\Ring^*$, and ask the challenger to sign a message $m$ w.r.t.\ $\Ring^*$ using the signing keys of member $i$.
\end{itemize}
 Notice that $\Ring^*$ may contain (potentially malicious) keys created by $\Adv$; but as long as the member $i$ is in both $\Ring^*$ and $\mathcal{R}$, the challenger is able to sign $m$ w.r.t.\ $\Ring^*$. The challenger also maintains a set $\mathcal{C}$, which records all the members in $\mathcal{R}$ that are corrupted by $\Adv$. To win the game, $\Adv$ needs to output a forgery $(\Ring^*, m^*, \Sigma^*)$ satisfying the following 3 requirements: 
 \begin{enumerate}
 \item
 $\Ring^* \subseteq \mathcal{R}\setminus \mathcal{C}$,
 \item
  $\RS.\Verify(\Ring^*,m^*, \Sigma^*) = 1$, {\bf and}
  \item
 $\Adv$ never made a signing query of the form $(\algo{sign}, \cdot, \mathcal{R}^*, m^*)$.
 \end{enumerate}

To consider quantum attacks, we first require that corruption queries should remain classical. In practice, corruption queries translate to the attack where a ring member is totally taken over by $\Adv$. Since ring signatures are a de-centralized primitive, corrupting a specific party should not affect other parties in the system. This situation arguably does not change with $\Adv$'s quantum power. One could of course consider ``corrupting a group of users in superposition'', but the motivation and practical implications of such corruptions is unclear, and thus we defer it to future research. In this work, we restrict ourselves to classical ring member corruptions.

We will allow $\Adv$ to conduct superposition attacks over the ring and message. That is, a QPT $\Adv$ can send singing queries of the form $(\algo{sign}, i, \sum_{\Ring,m} \psi_{\Ring,m}\ket{\Ring, m})$, where the identity $i$ is classical for the same reason above. Given the argument above, one may wonder why we allow superpositions over $\Ring$ in the signing query. The reason is that unlike for corruption queries, each signing query specifies a specific member $i$ to run the signing algorithm for. No matter what $\Ring$ is, this member will only sign using her own signing key (and this is the only signing key that she knows), and this has nothing to do with other parties in the system\footnote{Indeed,  $\Ring$ may even contain ``illegitimate'' or ``non-existent'' members faked by $\Adv$. Note that we do not require $\Ring\subseteq\mathcal{R}$.}. Therefore, superposition attacks over $\Ring$ can be validated just as superposition attacks over $m$, thus should be allowed. 

The next step is to determine the winning condition for QPT adversaries in the above quantum unforgeability game. The approach taken by \cite{C:CGHKLMPS21} is to extend the one-more unforgeability from \cite{C:BonZha13} to the ring setting.  Concretely, it is required that the adversary cannot produce $(\mathsf{sq} + 1)$ valid signatures by making only $\mathsf{sq}$ quantum {\sf sign} queries. However, there is a caveat. Recall that the $\Ring^*$ in $\Adv$'s forgery should be a subset of uncorrupted ring members (i.e., $\mathcal{R}\setminus\mathcal{C}$). A natural generalization of the ``one-more forgery'' approach here is to require that, with $\mathsf{sq}$ quantum signing queries, the adversary cannot produce $\mathsf{sq} + 1$ forgery signatures, where {\em all} the rings contained are subsets of $\mathcal{R}\setminus \mathcal{C}$. This requirement turns out to be so strict that, when restricted to the classical setting, this one-more unforgeability seems to be weaker than the standard unforgeability for ring signatures (more details in \Cref{sec:ring-sig:pq-definition} and \Cref{sec:one-more:PQ-EUF:ring-sig}).

Our idea is to extend the blind-unforgeability definition to our setting. Specifically, the challenger will create a blind set $B^\RS_\epsilon$ by including in each ring-message pair $(\Ring, m)$ with probability $\epsilon$. It will then blind the signing algorithm such that it always returns ``$\bot$'' for $(\Ring, m) \in B^\RS_\epsilon$. In contrast to one-more unforgeability, we will show that this definition, when restricted to the classical setting, is indeed equivalent to the standard unforgeability notion for ring signatures.

\para{Our Construction.} Our starting point is the LWE-based construction by Chatterjee et al.\ \cite{C:CGHKLMPS21}. 
We first recall their construction: the public key consists of a public key for a public-key encryption scheme $\PKE$ and a verification key for a standard signature scheme $\Sig$, as well as the first round message of a (bespoke) ZAP argument. To sign a message, one first computes an ordinary signature $\sigma$ and then encrypts this along with a hash key $hk$ for a specific hash function (i.e., {\em somewhere perfectly-binding} hash). Two such encryptions $(c_1,c_2)$ are produced, along with the second-round message $\pi$ of the ZAP proving that one of these encryptions is properly computed using a public key that is part of the presented ring. The hash key is extraneous to our concerns here; suffice it to say that it helps encode a ``hash'' of the ring into the signature and is a key feature in establishing compactness of their scheme.

To show anonymity, one starts with a signature for $i_0$, then switches the ciphertexts $c_1$ and $c_2$ in turn to be computed using the public key for $i_1$ while changing the ZAP accordingly. Semantic security ensures that ciphertexts with respect to different public keys are indistinguishable, and WI of the ZAP allows us to switch whichever ciphertext is not being used to prove $\pi$, and also to switch a proof for a ciphertext corresponding to $i_0$ to one corresponding to $i_1$. 

Unforgeability in \cite{C:CGHKLMPS21} follows from a reduction to the unforgeability of $\Sig$. Even though their construction uses a custom ZAP that only offers soundness for (effectively) $\NP \cap \coNP$, they develop techniques in this regard to show that even with this ZAP, one can ensure that if an adversary produces a forgery with non-negligible probability, then it also encrypts a valid signature for $\Sig$ in one of $c_1$ or $c_2$ with non-negligible probability. The reduction can extract this using a corresponding decryption key (which it can obtain during key generation for the experiment) and use this as a forgery for $\Sig$.

The \cite{C:CGHKLMPS21} construction can thus be seen as a compiler from ordinary to ring signatures assuming LWE. We use their template as a starting point, but there are significant differences between security notions for standard (classical) ring signatures, and our (quantum) blind-unforgeability setting. We discuss these and how to accomodate them next. The very first change that we require here is to use a blind-unforgeable signature scheme in lieu of $\Sig$, since we reduce unforgeability to that of $\Sig$. 

Next, let us discuss post-quantum anonymity. Here, the adversary can make a challenge query that contains a {\em superposition} over rings and messages. 
We would like to use the same approach as above, but of course computational indistiguishability is compromised against superposition queries. Two clear strengthenings are needed compared to the classical scheme: first, we need to use pairwise-independent hashing to generate signing randomness (to apply quantum oracle similarity techniques from \cite{C:BonZha13}). Second, we want to ensure statistical similarity of the components $c_1,c_2,\pi$ (in order to use an aforementioned lemma from \cite{C:BonZha13} which says that pointwise statistically close oracles are indistiguishable even with quantum queries). In particular, the $\PKE$ needs to be statistically close on different plaintexts, and the WI guarantee for the ZAP needs to be statistical. Fortunately, we can use lossy encryption for the constraint on ciphertexts, and the ZAP from \cite{C:CGHKLMPS21} is already statistical WI.   

Finally, we turn to blind-unforgeability. Here, the things that change are that firstly, we need to switch to injective public keys (instead of lossy ones) to carry over the reduction from the classical case. Further, we forego using SPB hashing, because our techniques require that we sign the message along with the ring, i.e. $\algo{Sig}.\Sign(sk, \Ring \| m)$. Thus we end up compromising compactness and using an SPB would serve no purpose. The reason that we need to sign the ring too has to do with how we define the blindset and how the challenger must maintain it in the course of the unforgeability game; this turns out to be more delicate than expected (see related discussion in \Cref{sec:discussion}). 
With the modifications above, we can eventually reduce the blind-unforgeability to that of $\Sig$. 
\section{Preliminaries}
{\bf Notation.} For a set $\mathcal{X}$, let $2^{\mathcal{X}}$ denote the power set of $\mathcal{X}$ (i.e., the set of all subsets of $\mathcal{X}$. Let $\secpar \in \Naturals$ denote the security parameter. 
A non-uniform QPT adversary is defined by $\Set{\mathsf{QC}_\secpar, \rho_\secpar}_{\secpar \in \Naturals}$, where $\Set{\mathsf{QC}_\secpar}_\secpar$  is a sequence of polynomial-size non-uniform  quantum circuits, and $\Set{\rho_\secpar}_\secpar$ is some polynomial-size sequence of mixed quantum states. For any function $F: \bits^n \rightarrow \bits^m$, ``quantum access'' will mean that each oracle
call to $F$ grants an invocation of the $(n + m)$-qubit unitary gate $\ket{x,t} \mapsto \ket{x, t \xor F(x)}$; we stipulate that for any $t \in \bits^*$, we have $t\xor\bot = \bot$. Symbols $\cind$, $\sind$ and $\idind$ are used to denote computational, statistical, and perfect indistinguishability respectively. Computational indistinguishability  in this work is by default w.r.t.\ non-uniform QPT adversaries.

We provide more preliminaries on lattices and the \cite{ITCS:BraVai14} key-homomorphic evaluation method in \Cref{sec:add-prelim}.

\subsection{Quantum Oracle Indistinguishability} 
We will need the following lemmata.
\begin{lemma}[\cite{C:Zhandry12}]\label{lemma:2qwise}
    Let $H$ be an oracle drawn from a 2q-wise independent distribution. Then, the advantage of any quantum algorithm making at most $q$ queries to $H$ has in distinguishing $H$ from a truly random function is 0. 
\end{lemma}
\begin{lemma}[\cite{C:BonZha13}]\label{lemma:oracleindist}
    Let $\mathcal{X}$ and $\mathcal{Y}$ be sets, and for each $x \in \mathcal{X}$, let $D_x$ and $D'_x$ be distributions on $\mathcal{Y}$ such that $|D_x-D'_x| \leq \epsilon$ for some value $\epsilon$ that is independent of $x$. Let $O: \mathcal{X} \rightarrow \mathcal{Y}$ be a function where, for each $x$, $O(x)$ is drawn from $D_x$, and let $O'(x)$ be a function where, for each $x$, $O'(x)$ is drawn from $D'(x)$. Then any quantum algorithm making at most $q$ queries to either $O$ or $O'$ cannot distinguish the two, except with probability at most $\sqrt{8C_0q^3\epsilon}$.
\end{lemma}

\subsection{Blind-Unforgeable Signatures}
We recall in \Cref{def:classical-BU-sig} the definition for blind unforgeable signature schemes in \cite{EC:AMRS20}.  The authors there provide a formal definition for MACs. We extend it in the natural way to the signature setting.
\begin{definition}[Blind-Unforgeable Signatures]\label{def:classical-BU-sig} 
For any security parameter $\secpar \in \Naturals$, let $\mathcal{M}_\secpar$ denote the message space and $\mathcal{T}_\secpar$ denote the signature space.
A {\em blind-unforgeable} signature scheme \Sig consists of the following PPT algorithms:
\begin{itemize}
\item 
$\Gen(1^\secpar)$ outputs a verification and signing key pair $(vk,sk)$. 

\item 
$\Sign(sk, m;r)$ takes as input a signing key $sk$, a message $m \in \mathcal{M}_\SecPar$, and a randomness $r$ (which we avoid specifying unless pertinent). It outputs a signature $\sigma \in \mathcal{T}_\secpar$.

\item 
$\Verify( vk, m, \sigma)$ takes as input a verification key $vk$, a message $m \in \mathcal{M}_\SecPar$ and a signature $\sigma \in \mathcal{T}_\SecPar$. It outputs a bit signifying accept (1) or reject (0).  
\end{itemize} 
These algorithms satisfy the following requirements: 
\begin{enumerate}
\item 
{\bf Completeness:} For any $\secpar \in \Naturals$, any $(vk, sk)$ in the range of $\Gen(1^\secpar)$, and any $m\in \mathcal{M}_\secpar$, it holds that $$\Pr[\Verify\big(vk,m,\Sign(sk,m)\big)=1]= 1-\negl(\secpar).$$

\item 
{\bf Blind-Unforgeability:} For any non-uniform QPT adversary $\Adv$, it holds w.r.t.\ \Cref{expr:BU:ordinary-signature} that 
		$$\algo{PQAdv}_\textsc{bu}^{\secpar}(\Adv) \coloneqq \Pr\big[\algo{PQExp}_\textsc{bu}^{\secpar}(\Adv) = 1\big] \le \negl(\secpar).$$
\begin{ExperimentBox}[
label={expr:BU:ordinary-signature},
]{Blind-Unforgeability Game \textnormal{$\algo{PQExp}^{\secpar}_\textsc{bu}(\Adv)$}}
\begin{enumerate}[label={\arabic*.}]
	\item 
	$\Adv$ sends a constant $0\le \epsilon \le 1$ to the challenger;

	\item 
	The challenger generates $(vk, sk) \leftarrow \Gen(1^\SecPar)$ and provides $vk$ to \Adv. 
	
	\item 
	The challenger defines a {\em blindset} $B^\Sig_\varepsilon \subseteq \mathcal{M}_\SecPar$ as follows: every $m \in \mathcal{M}_\SecPar$ is put in $B^\Sig_\varepsilon$ independently with probability $\varepsilon$. 

	\item
	$\Adv$ is allowed to make $\poly(\secpar)$ quantum queries. For each query, the challenger samples a (classical) random string $r$ and performs the following mapping:
	$$\sum_{m,t} \psi_{m,t}| m,t \rangle \mapsto \sum_{m,t} \psi_{m,t}| m,t \xor B^\Sig_\varepsilon\Sign(sk,m;r)\rangle,$$
	where $B^\Sig_\varepsilon\Sign(sk,m;r) =
					\begin{cases}
					\bot & \text{if}~m \in B^\Sig_\varepsilon \\
					\Sign(sk,m;r) & \text{otherwise}
					\end{cases}
					$.
	\item
	Finally, $\Adv$ outputs	$(m^*, \sigma^*)$; the challenger checks if:
	\begin{enumerate}
	 \item
	 $m^* \in B^\Sig_\epsilon$; {\bf and}	
	 \item
	 $\Verify(vk, m^*, \sigma^*) = 1$.
	\end{enumerate}
	If so, the experiment outputs 1; otherwise, it outputs 0.			

		
\end{enumerate}
\end{ExperimentBox}

\item {\bf Shortness (Optional):}The signature scheme is short if the signature size is at most a polynomial on the security parameter {\em and} the logarithm of the message size.
\end{enumerate}
\end{definition}


\begin{remark}[One randomness to rule them all\footnote{Inspired by J.\ R.\ R.\ Tolkien. Indeed, this is a ``ring'' signature paper.}]\label{rmk:single-randomness}
The signing algorithm in our definition samples signing randomness once per every query, as opposed to sampling signing randomness for every classical message in the superposition. This was established as a reasonable definitional choice in \cite{C:BonZha13}, where they observed that one could ``de-randomize'' the signing procedure by simply using a quantum PRF to generate randomness for each possible message in superposition, and use this for signing. 
We stick with this convention when defining post-quantum security for both ordinary signatures (\Cref{def:classical-BU-sig}) and ring signatures (\Cref{def:pq:anonymity,def:pq:blind-unforgeability}). 
\end{remark}
\begin{remark}
 We let the adversary choose $\varepsilon$. This is equivalent to quantifying over all values of $\varepsilon$ as in the definition in \cite{EC:AMRS20}. 
\end{remark}

\subsection{Quantum-Access Secure Biased Bit-PRF}
\label{sec:additional-prelims:biased-QPRF}

We will need a {\em quantum-access secure} PRF having a {\em biased single-bit} output. It should also be implementable by $\mathsf{NC}^1$ circuits. Let us first present the definition.
\begin{definition}[Biased Bit-QPRFs]\label{def:biased-bit-PRF} 
A biased bit-QPRF on domain $\bits^{n(\secpar)}$ consists of:
\begin{itemize}
\item
$\Gen(1^\secpar, \epsilon)$: takes as input a constant $\epsilon \in [0,1]$, outputs a key $k_\epsilon$;
\item
$\PRF_{k_\epsilon}(x)$: takes as input $x \in \bits^{n(\secpar)}$, outputs a bit $b \in \bits$,
\end{itemize}
such that for any $\epsilon\in[0,1]$ and any QPT $\Adv$ having {\em quantum access} to its oracle,
$$\big| \Pr\big[k_\epsilon \gets \Gen(1^\secpar, \epsilon): \Adv^{\PRF_{k_\epsilon}(\cdot)} = 1\big] - \Pr\big[F \pick \mathcal{F}\big(n(\secpar),\epsilon\big): \Adv^{F(\cdot)} = 1\big]\big|\le \negl(\secpar),$$
where $\mathcal{F}\big(n(\secpar),\epsilon\big)$ is the collection of all functions from $\bits^{n(\secpar)}$ to $\bits$ that output $1$ with probability $\epsilon$.
\end{definition}
It is known that the $\mathsf{NC}^1$ PRF from \cite{EC:BanPeiRos12} is quantum-access secure (i.e., a QPRF) \cite{FOCS:Zhandry12}. It can be made biased by standard techniques (e.g., using the standard QPRF to ``de-randomize'' a $\epsilon$-biased coin-tossing circuit). Note that the \cite{EC:BanPeiRos12} PRF relies on the quantum hardness of LWE with {\em super-polynomial} modulus. It is worthing noticing that such an LWE hardness assumption is stronger than the SIS assumption with
polynomial modulus (see \Cref{def:sis}).

\section{Blind-Unforgeable Signatures in the QROM}
\label{sec:BU:sig:QROM}

We show here that the signature scheme in \cite{STOC:GenPeiVai08} is a blind-unforgeable signature in the quantum random oracle model. This construction relies on the notion of {\em preimage sampleable functions}. 

\begin{definition}[Preimage Sampleable Functions \cite{STOC:GenPeiVai08}]
\label{def:PSF}
A {\em preimage sampleable function} family \PSF consists of the following PPT algorithms: 
\begin{itemize}
	\item $\Gen(1^\SecPar)$ samples a public/secret key pair $(pk,sk)$. 
\end{itemize}
For any $(pk, sk)$ in the range of $\Gen(1^\SecPar)$:
\begin{itemize}
	\item $\F(pk,\cdot)$ computes a function from set $\mathcal{X}_\SecPar$ to set $\mathcal{Y}_\SecPar$. 
	\item $\Samp(1^\SecPar)$ samples an $x$ from some (possibly non-uniform) distribution $\mathcal{X}_\SecPar$ such that $\F(pk,x)$ is distributed uniformly over $\mathcal{Y}_\SecPar$.
	\item $\InvF(sk,y)$ takes as input any $y \in \mathcal{Y}_\SecPar$ and outputs a preimage $x \in \mathcal{X}_\SecPar$ such that $\F(pk,x)=y$, {\em and} $x$ is distributed statistically close to $\Samp(1^\SecPar)$ conditioned on $\F(pk,x)=y$.
\end{itemize}
These algorithms satisfy the following properties:
\begin{enumerate}
\item \label{item:def:psf:preimage-min-entropy}
{\bf Preimage Min-entropy:} For each $y \in \mathcal{Y}_\SecPar$, the conditional min-entropy of $x \leftarrow \Samp(1^\SecPar)$ given $\F(pk,x)=y$ is at least $\omega(\log n)$.

\item  \label{item:def:psf:cr}
{\bf Collision Resistance:} For any QPT algorithm \Adv, the probability that $\Adv(1^\SecPar,pk)$ outputs distinct $x,x' \in \mathcal{X}_\SecPar$ such that $F(pk,x) = \F(pk,x')$ is negligible in $\SecPar$.
$$\Pr[
\begin{array}{l}
(pk, sk)\gets \Gen(1^\secpar);\\ (x, x')\gets \Adv(1^\SecPar,pk)
\end{array}: x\ne x' ~\wedge~ \F(pk,x) = \F(pk,x')] = \negl(\secpar).$$
\end{enumerate}
\end{definition}

\cite{STOC:GenPeiVai08} constructs such PSFs based on the hardness of the SIS problem. They also give a signature scheme using PSFs, a hash function modeled as a random oracle, and a pseudorandom function. In the following, we first recall their signature scheme in \Cref{prot:GPVSig}, and then prove in \Cref{thm:GPV:security-proof} that
this construction satisfies \Cref{def:classical-BU-sig} in the QROM if the PRF is a QPRF.

\begin{ConstructionBox}[label={prot:GPVSig}]{The GPV Signature \cite{STOC:GenPeiVai08}}
Let $\PSF$ be a preimage sampleable function. Let \PRF be a pseudorandom function, and $H$ be a hash function. The signature scheme \Sig is defined as follows: 

\subpara{$\Gen(\SecPar)$:}
	\begin{enumerate}
		\item Generate $(sk',pk') \leftarrow \PSF.\Gen(\SecPar)$;
		\item Sample a PRF key $k\gets \PRF.\Gen(1^\secpar)$;
		\item Output $sk=(sk',k)$ and $pk=pk'$.
	\end{enumerate}
\subpara{$\Sign(sk,m)$:} 
	\begin{enumerate}
		\item Compute $r \leftarrow \PRF(k,m)$ and $h=H(m)$;
		\item Compute $\sigma = \InvF(sk',h;r)$;
		\item Output $\sigma$.
	\end{enumerate}
\subpara{$\Verify(pk,m,\sigma)$:} 
	\begin{enumerate}
		\item Compute $h=H(m)$ and $h'=\F(pk',\sigma)$;
		\item If $h=h'$ output $1$; otherwise, output $0$.
	\end{enumerate} 
\end{ConstructionBox}

\begin{theorem}\label{thm:GPV:security-proof}
Assume that \PSF be a preimage sampleable function, PRF is a quantum secure pseudorandom function, and $H$ realizes a random oracle. Then \Cref{prot:GPVSig} is a short blind-unforgeable signature in the quantum random oracle model.
\end{theorem}

\begin{proof} 
Completeness and shortness of \Cref{prot:GPVSig} are straightforward. In the following, we prove blind-unforgeability.

\para{The Joint Oracle.} First notice that in the blind-unforgeability game in the ORAM, the adversaries has quantum oracle access to {\em two} oracles: $H$ and $B_\varepsilon\Sign$. As we will show later (in particular, in hybrids $H_1$ and $H_2$ below), we need to change the way these two oracles are sampled, without being noticed by the adversary. We will need to argue the indistinguishability of this switch by invoking \Cref{lemma:oracleindist}; but \Cref{lemma:oracleindist} is for adversaries that have access  to  a {\em single} oracle ($\mathcal{O}$ or $\mathcal{O}'$). Therefore, we start by slightly changing our oracle interface. 

Instead of maintaining separate random and signing oracles (i.e., $H$ and $B_\varepsilon\Sign$ respectively), we will maintain a single {\em joint oracle} $\mathcal{O}$ that can answer both types of queries `jointly'. In more detail, we ask the adversary to include a {\em flag} bit $c \in \bits$ in each query. If $c = 0$,  $\mathcal{O}$ will respond as $H$; if $c = 1$, it will respond as $B_\epsilon\Sign$. Formally, $\mathcal{O}$ implements the following mapping:
$$\sum_{c,m,t}\Psi_{c,m,t} \ket{c,m,t} \mapsto \sum_{c,m,t}\Psi_{c,m,t} \ket{c,m,t \xor G(c, m)},$$
where $G(c,m) = 
\begin{cases}
H(m) & c = 0 \\
B_\epsilon \Sign(m) & c =1
\end{cases}$.



This transformation is without loss of generality---indeed, any adversary $\Adv'$ that wins the original blind-unforgeability game can be transformed into an adversary \Adv that breaks blind-unforgeability w.r.t.\ the joint oracle $\mathcal{O}$. \Adv need only forward queries from $A'$ to $\mathcal{O}$ and corresponding responses back to $\Adv'$ by setting a proper (classical) bit $c$ . 
It is straightforward to see that $\Adv'$ gets identical responses whether interacting with \Adv or in the QROM challenge. We therefore conclude that \Adv has the same success probability as $\Adv'$, and this validates our single joint oracle interface. For the rest of this proof, we presuppose an adversary \Adv that directly interacts with the joint oracle. 

We will use \Adv to obtain a contradiction, by employing hybrid arguments. For ease of exposition we will also use the following shorthand: denoting $\F(pk,\cdot)$ by $f(\cdot)$, and $\InvF(sk,\cdot)$ by $f^{-1}(\cdot)$. Consider the following hybrid experiments: 

\para{Hybrid $H_0$:} This is simply the normal blind-unforgeability challenge with the joint oracle. In particular, for each $m$ in the superpostion of the adversary's quantum query $\Sigma_m \Psi_m \ket{m}$, the hash $h$ is computed (implicitly) as the random oracle output $H(m)$, and the signature $\sigma_m$ is computed according to $\sigma_m = f^{-1}(h;r)$ where $r=\PRF_k(m)$ (note that, in accordance with the challenge, the signing algorithm will be invoked only if $m \notin B_\varepsilon$).  

\para{Hybrid $H_1$:} In this hybrid we change how $r$ is generated. Instead of computing $r=\PRF_k(m)$, we set $r=J(m)$ where $J(\cdot)$ is a {\em random} function over the range and domain of $\PRF$.\footnote{Note that the adversary's query is quantum: $\Sigma_m \Psi_m \ket{m}$. To keep the notation succinct, it suffices to describe the computation for each $m$ in the superposition. We stick to this convention for later hybrids in this proof.}  

\subpara{$\Output(H_0) \cind \Output(H_1)$:} This follows directly from the quantum security of the PRF. Any adversary that has distinguishable outputs in these two hybrids is easily converted into a QPT algorithm that can distinguish between the QPRF and a uniformly random function given quantum oracle access to them in turn. 

\para{Hybrid $H_2$:} In this hybrid we change both components of the oracle. The sining oracle is now re-defined as $\sigma_m = \Sig.\Sign(m) \coloneqq \Samp\big(1^\SecPar;J(m)\big)$. Further, the hash oracle query is now answered by $h=f(\sigma_m)$. That is, when the adversary asks for $H(m)$, the hybrid first computes $\sigma_m = \Samp\big(1^\SecPar; J(m)\big)$, and then returns $h = f(\sigma_m)$ to the adversary.   

\subpara{$\Output(H_1) \sind \Output(H_2)$:} Recall that we view the random oracle $H$ and the signing oracle together as a joint oracle. The only thing that changes in $H_2$ is the computation of parts of the joint oracle response. We go through these carefully. For every $m$ (in the superposition of $\Adv$'s quantum query), the response changes from 
\begin{itemize}
\item
$\Big(H(m),~f^{-1}\big(H(m);J(m)\big)\Big)$, i.e., the joint oracle in $H_1$, {\bf to}
\item
$\Big(f\big(\Samp(\SecPar;J(m))\big),~\Samp\big(\SecPar;J(m)\big)\Big)$, i.e, the joint oracle in $H_2$.
\end{itemize} 
Note that both $H(m)$ and $J(m)$ are uniformly random. Therefore, by the properties of $\Samp$ and $f^{-1}$ (\Cref{def:PSF}), the above two distributions are statistically close to each other. Denoting the joint oracle in $H_1$ by $\mathcal{O}_1$, and that in $H_2$ by $\mathcal{O}_2$, we conclude that for any (classical) query point $m$, the distributions of the responses returned by $\mathcal{O}_1$ and $\mathcal{O}_2$ conditioned on $m$ are statistically close, say less than distance $\Delta(\SecPar)$ (which is negligible in $\SecPar$). Now since \Adv is a quantum machine making at most polynomially (say $q(\secpar)$) many quantum queries. Then, we can use \Cref{lemma:oracleindist} to conclude that \Adv distinguishes between $\mathcal{O}_1$ and $\mathcal{O}_2$ with probability at most $\sqrt{8C_0q^3\Delta}$, which is negligible in $\SecPar$. 

\para{Hybrid $H_3$:} Observe that the hybrid $H_2$ is {\em not} efficiently implementable as it needs to sample a random function $J(\cdot)$. In the classical setting, $H_2$ can be made efficient by lazy-sampling $J(\cdot)$; however, here we cannot resort to lazy-sampling as the adversary has {\em quantum} access to the oracle. Thus, we take the following alternative approach to have an efficient hybrid. Assume $q$ is the upper-bound of the number of quantum queries made by the adversary. In hybrid $H_3$, we sample a $2q$-wise independent hash function $J'(\cdot)$, and replace the random function $J(\cdot)$ with $J'(\cdot)$. Everything else remains the same as in $H_2$. 

\subpara{$\Output(H_2) \idind \Output(H_3)$}: This follows immediately from \Cref{lemma:2qwise}.

\para{Reducing to the security of PSF.} Now consider the eventual forgery output by \Adv in $H_3$, $(m^*,\sigma_m^*)$. If this is valid, it follows from \Cref{def:classical-BU-sig} that $m^*\in B_\epsilon$ and $\Sig.\Verify(vk, m^*, \sigma_m^*) = 1$, which means $H(m^*)=f(\sigma_m^*)$.

Let $\sigma_m' \coloneqq \Samp(1^\SecPar;J'(m^*))$. Note that this is exactly $\Sig.\Sign$'s output on $m^*$ in $H_3$. Due to the way $H_3$ implements the oracles, this presents only two possibilities:
 \begin{enumerate}
 \item 
  Either we have $\sigma_m' = \sigma_m^*$, in which \Adv is able to pick out the value $\sigma_m'$ among all possible preimages of $h= f(\sigma_m^*)$. Observe that $m^* \in B_\epsilon$, which means that $\Adv$ never saw that value $\sigma_m'$ before as $\Sig.\Sign$ returns $\top$ for messages in $B_\epsilon$.
  Therefore, by the preimage min-entropy property (\Cref{item:def:psf:preimage-min-entropy}) of PSF, the set of allowed pre-images for $h$ has conditional minentropy at least $\omega(\log \SecPar)$, which means that \Adv can only predict this value with at most negligible probability, so this case has a negligible chance of occurrence. 
  \item
  Else, we must have $\sigma_m' \neq \sigma_m^*$ in which case \Adv has obtained colliding preimages, i.e. $\sigma_m' \neq \sigma_m^*$ such that $f(\sigma_m') = f(\sigma_m^*)$. This contradicts the collision resistance of PSF (note that this is the reason why we need $H_3$ to be efficiently implementable), so this case can also only arise with at most negligible probability.
 \end{enumerate} 
 We conclude that a successful forgery occurs only with negligible probability if the PSF satisfies the aforementioned properties.  
\end{proof}
\section{Blind-Unforgeable Signatures in the Plain Model}
\subsection{Notation and Building Blocks}
 We assume familiarity with standard lattice-based cryptographic notions and procedures. Here we will recall certain techniques and properties to be directly used in our plain model construction.
We recall standard lattice-related concepts (e.g., parameters, hardness, trapdoors) in \Cref{sec:add-prelim:lattice}. 

For a vector $\mathbf{u}$, we let $||\mathbf{u}||$ denote its $\ell_2$ norm. For a matrix $\mathbf{R} \in \mathbb{Z}^{k\times m}$, we define two matrix norms: 
\begin{itemize}
\item
$||\mathbf{R}||$ denotes the $\ell_2$ norm of the largest column of $\mathbf{R}$;
\item
 $||\mathbf{R}||_2$ denotes the operator norm of $\mathbf{R}$, defined as $||\mathbf{R}||_2 = \mathsf{sup}_{\vb{x}\in \mathbb{R}^{m+1}}||\mathbf{R}\cdot\mathbf{x}||$.
\end{itemize} 
We denote the Gram-Schmidt ordered orthogonalization of a matrix $\vb{A} \in \mathbb{Z}^{m\times m}$ by $\tilde{\vb{A}}$. For a prime $q$, a modular matrix $\mathbf{A} \in \mathbb{Z}_q^{n\times m}$ and vector $\mathbf{u} \in \mathbb{Z}_q^n$, we define the m-dimensional (full rank) lattice $\Lambda_q^{\vb{u}}(\mathbf{A}) = \Set{\mathbf{e} \in \mathbb{Z}^m:\mathbf{Ae}=\vb{u}~(\mathrm{mod}~q)}$. In particular, $\Lambda_q^\perp(\mathbf{A})$ denotes the lattice $\Lambda_q^{\vb{0}}(\mathbf{A})$.


\para{Lattice Sampling Algorithms.} Our construction uses the `left-right trapdoors' framework introduced in \cite{EC:AgrBonBoy10,PKC:Boyen10}, which uses two sampling algorithms $\algo{SampleLeft}$ and $\algo{SampleRight}$. 

\subpara{$\algo{SampleLeft}$:} The algorithm $\algo{SampleLeft}$ works as follows: 
\begin{itemize}
\item {\em Inputs:} A full-rank matrix $\mathbf{A} \in \mathbb{Z}_q^{n\times m}$ and a short basis $\mathbf{T_A}$ of $\Lambda_q^\perp(\mathbf{A})$, along with a matrix $\mathbf{B} \in \mathbb{Z}_q^{n\times m_1}$, a vector $\mathbf{u} \in \mathbb{Z}_q^{n}$, and a Gaussian parameter $s$. 
\item {\em Output:} Let $\mathbf{F} = [\mathbf{A}~|~\mathbf{B}]$. $\algo{SampleLeft}$ outputs a vector $\mathbf{d} \in \mathbb{Z}^{m+m_1}$ in $\Lambda_q^\mathbf{u}(\mathbf{F})$. 
\end{itemize}

\begin{theorem}[$\algo{SampleLeft}$ Closeness \cite{EC:AgrBonBoy10,EC:CHKP10}]\label{theorem:sampleleft}
Let $q > 2$, $m > n$ and $s > ||\widetilde{\vb{T}}_{\vb{A}}||\cdot\omega\big(\sqrt{\log (m+m_1)}\big)$. Then, the $\algo{SampleLeft}(\mathbf{A},\mathbf{B},\mathbf{T_A},\mathbf{u},s)$ (as defined above) outputs $\mathbf{d} \in \mathbb{Z}^{m+m_1}$ distributed statistically close to $\mathcal{D}_{\Lambda_q^\mathbf{u}(\mathbf{F}),s}$. 
\end{theorem}

\subpara{$\algo{SampleRight}$:} The algorithm $\algo{SampleRight}$ works as follows: 
\begin{itemize}
\item {\em Inputs:} Matrices $\mathbf{A} \in \mathbb{Z}_q^{n\times k}$ and $\mathbf{R} \in \mathbb{Z}_q^{k\times m}$, a full-rank matrix $\mathbf{B} \in \mathbb{Z}_q^{n\times m}$, a short basis $\mathbf{T_B}$ of $\Lambda_q^\perp(\mathbf{B})$, a vector $\mathbf{u} \in \mathbb{Z}_q^{n}$, and a Gaussian parameter $s$.  
\item {\em Output:} Let $\mathbf{F} = [\mathbf{A}~|~\mathbf{AR}+\mathbf{B}]$. It outputs a vector $\mathbf{d} \in \mathbb{Z}^{m+m_1}$ in the set $\Lambda_q^\mathbf{u}(\mathbf{F})$. 
\end{itemize}

\begin{theorem}[$\algo{SampleRight}$ Closeness \cite{EC:AgrBonBoy10}]\label{theorem:sampleright}
Let $q > 2$, $m > n$ and $s > ||\widetilde{\vb{T}}_{\vb{B}}||\cdot ||\vb{R}||_2 \cdot \omega(\sqrt{\log m})$. Then $\algo{SampleRight}(\mathbf{A},\mathbf{B},\mathbf{R},\mathbf{T_B},\mathbf{u},s)$ (as defined above) outputs $\mathbf{d} \in \mathbb{Z}^{m+k}$ distributed statistically close to $\mathcal{D}_{\Lambda_q^\mathbf{u}(\mathbf{F}),s}$. 
\end{theorem}

\para{Random Sampling Related.}
The following is a simple corollary of \cite[Lemma 4]{EC:AgrBonBoy10} (see \Cref{sec:additional-prelims:sampling-related} for details). 

\begin{corollary}\label{lemma:ext-lhl}
 Suppose that $m > (n+1)\log_2q+\omega(\log n)$ and that $q > 2$ is a prime. Let $\mathbf{R}$ be an $m\times k$ matrix chosen uniformly from $\Set{-1,1}^{m\times k}\mod q$ where $k=k(n)$ is polynomial in $n$. Let $\mathbf{A}' \in \mathbb{Z}_q^{n\times m}$ be sampled from a distribution statistically close to uniform over $\mathbb{Z}_q^{n\times m}$. Let $\mathbf{R}$ be an $m\times k$ matrix chosen uniformly from $\Set{-1,1}^{m\times k}\mod q$ where $k=k(n)$ is polynomial in $n$. Let $\mathbf{B}$ be chosen uniformly in $\mathbb{Z}_q^{n\times k}$. Then for all vectors $\mathbf{w} \in \mathbb{Z}_q^m$, the distributions $(\mathbf{A}',\mathbf{A}'\vb{R},\mathbf{R}^\top\mathbf{w})$ and $(\mathbf{A}',\mathbf{B},\mathbf{R}^\top\mathbf{w})$ are statistically close. 
\end{corollary}

\para{Key-Homomorphic Evaluation.}
We briefly recall the matrix key-homomorphic evaluation algorithm, as found in \cite{C:GenSahWat13,EC:BGGHNS14,ITCS:BraVai14} (see \Cref{sec:additional-prelims:key-homo-eval} for details). This template evaluates NAND circuits, gate by gate, in a homomorphic manner. For a {\sf NAND} gate $g(u,v;w)$ with input wires $u,v$ and output wire $w$, we have (inductively) matrices $\mathbf{A}_u = \mathbf{AR}_u + x_u\mathbf{G}$, and $\mathbf{A}_v = \mathbf{AR}_v + x_v\mathbf{G}$ where $x_u$ and $x_v$ are the input bits of $u$ and $v$, and the evaluation algorithm computes:
$$\mathbf{A}_w 
  = \mathbf{G} - \mathbf{A_u}\cdot\mathbf{G}^{-1}(\mathbf{A}_v) 
= \mathbf{G} - (\mathbf{AR}_u+x_u\mathbf{G})\cdot\mathbf{G}^{-1}(\mathbf{AR}_v + x_v\mathbf{G}) 
 = \mathbf{AR}_g + (1-x_ux_v)\mathbf{G},
$$
where $1-x_ux_v \coloneqq \algo{NAND}(x_u,x_v)$, and $\mathbf{R}_g = -\mathbf{R}_u\cdot\mathbf{G}^{-1}(\mathbf{A}_v)-x_u\mathbf{R}_v$ has low norm if both $\mathbf{R}_u$ and $\mathbf{R}_v$ have low norm.

\para{Biased Bit-QPRF}. We will need a single-bit-output QPRF that output 1 with a customizable probability $\epsilon$. Moreover, we need it to be implementable in $\mathsf{NC}^1$. Such a QPRF can be built using the PRF constructed in \cite{EC:BanPeiRos12} assuming QLWE with super-polynomial modulus. See \Cref{sec:additional-prelims:biased-QPRF} for more details and the formal definition.

\subsection{Our Construction}
Our signature scheme uses a biased bit QPRF $\PRF$ whose input space $\mathcal{X}$ corresponds to our message space $\mathcal{M}$, and the algorithms $\algo{SampleLeft}$, $\algo{SampleRight}$ given as in \Cref{theorem:sampleleft} and \Cref{theorem:sampleright} respectively, and $\algo{TrapGen}$ that can sample matrices in $\mathbb{Z}_q^{n\times m}$ statistically close to uniform, along with a corresponding `short' or `trapdoor' basis for the associated lattice. This is formally defined in \Cref{lemma:trapgen}. The construction is as follows:
\begin{ConstructionBox}[label={constr:BU:plain}]{Blind-Unforgeable Signatures in the Plain Model}
{\bf Paramters:} Set message length $t(\secpar)$ and row size $n(\secpar)$ as free parameters (polynomial in $\secpar$). PRF key size is set as $k(\secpar)$, and the depth for $\mathsf{C}_{\algo{PRF}}$ is given by $d(\secpar)$. We set $m=n^{1+\eta}$ for proper running of $\algo{TrapGen}$, and $\mathsf{sigsize}_\secpar = s\sqrt{2m}$ for the validity of $\algo{SampleLeft}$ output (to ensure completeness). Set $s = O(4^dm^{3/2})\cdot\omega(\sqrt{\log m})$ to ensure statistical closeness of $\algo{SampleLeft}$ and $\algo{SampleRight}$, and correspondingly set {$\beta = O(16^dm^{7/2})\cdot\omega(\sqrt{\log m})$} and {$q = O(16^dm^4)\cdot \big(\omega(\sqrt{\log m})\big)^2$} to have an overall reduction to an appropriately hard instance of SIS. For further details about these choices, see \Cref{sec:appendix:parameters-plain-BU-sig}. 

\subpara{$\Gen(1^\SecPar)$:}
	\begin{enumerate}
		\item 
		Sample a matrix $\mathbf{A}$ along with a `trapdoor' basis $\mathbf{T_A}$ for $\Lambda^\perp_q(\mathbf{A})$ using $\algo{TrapGen}$.
		\item 
		Sample a matrix $\mathbf{A}'$, `PRF key' matrices $\mathbf{B}_1,\dots,\mathbf{B}_k$, and `PRF input' matrices $\mathbf{C}_0,\mathbf{C}_1$ uniformly from $\mathbb{Z}_q^{n \times m}$ ($k$ is the PRF key length).
		\item 
		Fix the Gaussian width parameter $s$ as given in parameter selection. 
		\item Fix a Boolean circuit description $\mathsf{C}_{\algo{PRF}}$ of the algorithm $\algo{PRF}_{(\cdot)}(\cdot)$. 
		\item Output $vk=(\mathbf{A},\mathbf{A}',\Set{\mathbf{B}_i}_{i=1}^k, \{\mathbf{C}_0,\mathbf{C}_1\}, \algo{PRF},s,\mathbf{C}_{\PRF})$ and $sk=\mathbf{T_A}$. 
	\end{enumerate}
\subpara{$\Sign(sk,vk,M)$:} let $(M_1,\dots,M_t) \in \bits^t$ be the bit-wise representation of $M$.
	\begin{enumerate}
		\item Run the \cite{ITCS:BraVai14} evaluation algorithm $\Eval_\textsc{bv}$ to homomorphically evaluate the circuit $\mathbf{C}_{\PRF}$ using the `encoded' PRF key bits $\Set{\mathbf{B}_i}_{i\in[k]}$ and message bits $\Set{\mathbf{C}_{M_j}}_{j \in[t]}$. This yields 
		$$\mathbf{A}_\textsc{prf,m} \coloneqq \Eval_\textsc{bv}(\mathbf{C}_{\PRF},\Set{\mathbf{B}_i}_{i\in[k]},\Set{\mathbf{C}_{M_j}}_{j \in [t]}) \in \mathbb{Z}_q^{n \times m}.$$
		\item 
		Set $\mathbf{F}_\textsc{m} \coloneqq [\mathbf{A}~|~\mathbf{A}' - \mathbf{A}_\textsc{prf,m}]$.
		\item
		Use $\algo{SampleLeft}$ to obtain $\mathbf{d}_\textsc{m}$ with distribution statistically close to $\gets \mathcal{D}_{\Lambda_q^\perp(\mathbf{F}_\textsc{m}),s}$ (see \Cref{theorem:sampleleft}).
		\item 
		Output $\sigma = \mathbf{d}_\textsc{m} \in \mathbb{Z}_q^{2m}$.
	\end{enumerate}
\subpara{$\Verify(vk,M,\sigma)$:} 
	\begin{enumerate}
		\item 
		Compute $\mathbf{A}_\textsc{prf,m}$, $\mathbf{F}_\textsc{m}$ as before.
		\item 
		Check that $\sigma \in \mathbb{Z}_q^{2m}$, $\sigma\neq 0$, and $||\sigma|| \leq \mathsf{sigsize}_\secpar$. If it fails, output $0$.  
		\item 
		If $\mathbf{F}_\textsc{m} \cdot \sigma = 0 \mod q$, output $1$, otherwise output $0$. 
	\end{enumerate}
\end{ConstructionBox} 

\subsection{Parameter Selection for \Cref{constr:BU:plain}}
\label{sec:appendix:parameters-plain-BU-sig}
The security parameter $\secpar$ is represented as before. We set message length $t(\secpar)$ and row size $n(\secpar)$ as free parameters (polynomial in $\secpar$). PRF key size is set as $k(\secpar)$, and the depth for $\mathsf{C}_{\algo{PRF}}$ is set to be $d(\secpar)$. We must set the parameters properly to ensure that the following conditions are satisfied:
\begin{enumerate}
	\item We must have $m=n^{1+\eta}$, with $\eta$ being given by $n^\eta > O(\log q)$. This is to ensure that $\algo{TrapGen}$ can run properly, by \Cref{lemma:trapgen}.
	\item We require that $s > ||\widetilde{\mathbf{T}}_{\vb{G}}||\cdot||\mathbf{R}||_2\cdot\omega(\sqrt{\log m})$, where $\mathbf{R} = \mathbf{R_{A'}} - \mathbf{R}_{\mathbf{A}_{\mathrm{PRF},\mathrm{M}}}$ (the latter will be defined in the course of the proof), for the statistical similarity of $\algo{SampleLeft}$ and $\algo{SampleRight}$, as per \Cref{theorem:sampleleft,theorem:sampleright}.
	\item Since signatures are vectors of length $2m$ over $\mathbb{Z}_q$ sampled from (statistically close to) $\mathcal{D}_{\Lambda_q^\perp(\mathbf{F}_m),s}$, for most honestly generated signatures to be valid, it is necessary to set $\mathsf{sigsize}_\secpar \geq s\sqrt{2m}$, in accordance with \Cref{lemma:discgaussconc}. 
	\item For hardness of the SIS instance, we require that the width parameter {$\beta$ satisfy $\beta \geq O(4^d\cdot m^{3/2}\cdot s\sqrt{2m})$}.
	\item Finally, for standard average-to-worst case hardness reductions to apply for SIS, we require that {$q \geq \beta \cdot\omega(\sqrt{n\log n})$}.  
\end{enumerate} 

Accordingly, we set the remaining parameters as follows: 
\begin{itemize}
	\item We set $m=n^{1+\eta},\mathsf{sigsize}_\secpar=s\sqrt{2m}$ just as indicated. 
	\item By the bounds on $\mathbf{R}_{\mathbf{A}_{\mathrm{PRF},\mathrm{M}}}$ implied by \Cref{lemma:rnorm}, it suffices to set $s=O(4^dm^{3/2})\cdot\omega(\sqrt{\log m})$ to satisfy constraint 2 above. 
	\item Using the above $s$ and so as to just about satisfy constraint 4, we set {$\beta = O(16^dm^{7/2})\cdot\omega(\sqrt{\log m})$}. 
	\item We set $q$ based on $\beta$ above so as to just satisfy the final constraint, namely {$q = O(16^dm^4)\cdot\omega(\sqrt{\log m})^2$}. 
\end{itemize}

Since we consider PRFs in $\mathsf{NC}^1$, we can write $d = c \log \ell$ (for some constant $c$) where $\ell = t+k$ is the input length for the PRF. This yields {$\beta = O(\ell^{4c}m^{7/2})\cdot\omega(\sqrt{\log m})$ and $q = O(\ell^{4c}m^4)\cdot\omega(\sqrt{\log m})^2$}. 

\subsection{Proof of Security} 
Completeness follows straightforwardly from the correctness of $\algo{SampleLeft}$ (\Cref{theorem:sampleleft}) for $\mathcal{D}_{\Lambda_q^\bot(\mathbf{F}),s}$. In the following, we prove BU-security.
\begin{theorem}\label{thm:plain-BU-sig:security}
Let $\secpar$ denote the security parameter, and $\algo{PRF}$ be a biased bit QPRF as defined in \Cref{def:biased-bit-PRF} above. If the parameters $n,m,q,\beta,s,d$ are picked as discussed above, and the $\mathbf{SIS}_{q,\beta,n,m}$ problem is hard for QPT adversaries, then our signature scheme $\Sig$ constructed as above, with the indicated parameters, satisfies Blind-Unforgeability as in \Cref{def:pq:blind-unforgeability}. 
\end{theorem}

\begin{proof}
Consider a QPT $\Adv$ that is able to produce forgeries w.r.t.\ $\Sig$ in the blind-unforgeability challenge. Our proof proceeds using a series of hybrid experiments. In the final hybrid we show a reduction from an adversary producing succesful forgeries to the hardness of $\mathbf{SIS}_{q,\beta,n,m}$. The hybrids are as follows: 

\para{Hybrid $H_0$:} This is the blind-unforgeability game (\Cref{expr:BU:ordinary-signature}). Namely, for an adversary-specified $\epsilon$, the challenger manually samples an $\epsilon$-weight set $B_\epsilon$ over messages, and does not answer queries in $B_\epsilon$. Signing and verification keys are chosen just as in the ordinary signing procedure. 

\para{Hybrid $H_1$:} This hybrid is identical to the previous one, except that we change the ordinary key generation into the following: 
\begin{enumerate}
	\item 
	Sample $\mathbf{A}$  with a `trapdoor' basis $\mathbf{T_A}$ for $\Lambda^\perp_q(\mathbf{A})$ using $\algo{TrapGen}$ as before. 
	\item  
	Sample `low-norm' matrices: $\mathbf{R_{A}'},\Set{\mathbf{R}_{\vb{B}_i}}_{i=1}^k,\mathbf{R_C}_0,\mathbf{R_C}_1 \pick \Set{-1,1}^{m\times m}$.
	\item 
	Let $\algo{PRF}$ and $\mathsf{C}_{\algo{PRF}}$ be as before. 
	\item \label[Step]{item:PQSig:plain:h-1:4}
	Sample a PRF key $k_\epsilon \gets \algo{PRF.Gen}(1^\secpar,\epsilon)$, where $k_\epsilon = s_1,\dots,s_k$ (i.e. has length $k$).
	\item 
	Set $\mathbf{A}' = \mathbf{AR}_{\vb{A}'} + \mathbf{G}$, where $\vb{G}$ the gadget matrix $\mathbf{G}$, which has a publicly-known trapdoor $\widetilde{\mathbf{T}}_{\mathbf{G}}$ (as described in \Cref{lemma:gadgettrapdoor}). 
	\item \label[Step]{item:PQSig:plain:h-1:6}
	Set $\mathbf{C}_b = \mathbf{AR}_{\vb{C}_b} + b\mathbf{G}$ for $b \in \Set{0,1}$, and sample $\mathbf{B}_i \pick \mathbb{Z}_q^{n \times m}$ for every $i \in [k]$.  
	\item 
	Fix the Gaussian width parameter $s$ as before.
	\item 
	Output $vk=(\mathbf{A},\mathbf{A}',\Set{\mathbf{B}_i}_{i=1}^k, \{\mathbf{C}_0,\mathbf{C}_1\}, s, \algo{PRF},\mathbf{C}_{\PRF})$, and $sk=(\mathbf{T_A},k_\epsilon)$.
\end{enumerate}
Note that while this hybrid generates a key $k_\epsilon$, it never uses it. 

\subpara{$H_0 \sind H_1$:} The only thing that changes (w.r.t.\ $\Adv$) is the distribution of the various components $(\mathbf{A'},\mathbf{C}_0,\mathbf{C}_1)$ of the verification key handed out by the challenger. However, by \Cref{lemma:ext-lhl} these distributions are all statistically close to the corresponding distributions in $H_0$. Note that the verification key is picked at the start of the challenge and provided to $\Adv$, so there is no scope for $\Adv$ to have quantum access to these component distributions. Thus the outputs in these hybrids are statistically close.

\para{Hybrid $H_2$:} This hybrid is identical to the previous one, except that we change how the challenger picks the blindset---Instead of manually sampling $B_\epsilon$ as a random $\epsilon$-weight set, it now sets $B_\epsilon$ to be the set of messages $M$ where $\algo{PRF}_{k_\epsilon}(M)$ is 1 (note that the challenger now possesses $k_\epsilon$ as part of $sk$, and can compute $\algo{PRF}_{k_\epsilon}(\cdot)$). Observe that the challenger in this hybrid is now efficient. 

\subpara{$H_1 \cind H_2$:} Note that setup and key generation in $H_2$ is identical to that in $H_1$---In particular, the adversary learns {\em no} information about the key $k_\epsilon$. The indistinguishability between $H_1$ and $H_2$ then follows immediately from the security of the biased bit-QPRF (\Cref{def:biased-bit-PRF}).

\para{Hybrid $H_3$:}  This hybrid is identical to the previous one, except that we change how the matrices $\vb{B}_i$'s (in \Cref{item:PQSig:plain:h-1:6}) are generated. Namely, we now set $$\forall i \in [k], ~~\mathbf{B}_i \coloneqq \mathbf{AR}_{\vb{B}_i} + s_i \cdot\mathbf{G}.$$ 
(Recall that $s_i$ is the $i$-th bit of the $k_\epsilon$ generated in \Cref{item:PQSig:plain:h-1:4}.)

\subpara{$H_2 \sind H_3$:} The only things that change between these hybrids are the matrices $\Set{\vb{B}_i}_{i \in [k]}$. Again, using \Cref{lemma:ext-lhl} the distributions for $\vb{B}_i$ for each $i \in [k]$ are all statistically close to the corresponding distributions in $H_2$, and just as in the similarity argument between $H_2$ and $H_3$, we can conclude that these hybrids too have indistinguishable outputs.

\para{Hybrid $H_4$:} Observe that, starting from $H_1$, we have: 
\begin{align*}
\mathbf{F}_\textsc{m} 
& = [\vb{A} ~|~ \vb{A}' - \vb{A}_\textsc{prf,m}] = 
\big[\vb{A} ~|~ \vb{A}' - \Eval_\textsc{bv}(\mathbf{C}_{\PRF},\Set{\mathbf{B}_i}_{i\in[k]},\Set{\mathbf{C}_{M_j}}_{j \in [t]}) \big]\\
& =  \big[ \vb{A} ~|~ \vb{A}' - \big( \vb{A}\vb{R}_\textsc{prf,m} + \PRF_{k_\epsilon}(M) \cdot\vb{G} \big) \big]\\
& = \big[\mathbf{A}~|~\mathbf{A}(\mathbf{R_{A'}}-\mathbf{R}_\textsc{prf,m}) + \big(1-\algo{PRF}_{k_\epsilon}(M)\big)\cdot\mathbf{G}\big].
\end{align*}
In this hybrid, we switch to using $\algo{SampleRight}$ to answer signing queries, instead of using $\algo{SampleLeft}$.  That is, we run $\algo{SampleRight}$ using $\mathbf{T_G}$, the publicly available trapdoor for $\mathbf{G}$. Note this means that now the challenger cannot answer queries where the `right half' of $\mathbf{F}_\textsc{m}$ does not include $\mathbf{G}$, i.e., $\algo{PRF}_{k_\epsilon}(M) =1$. But due to the way $H_2$ generate the blindset, such a query is anyway answered with ``$\bot$''.

\subpara{$H_3 \cind H_4$:} We first show that these two hybrids answer signature queries for any {\em classical} query $M$ in a {\em statistically} indistinguishable manner. For any query $M$, there are two cases: (1) if $\algo{PRF}_{k_\epsilon}(M) =1$, the challengers in both $H_3$ and $H_4$ return $\bot$. In this case, these distributions are identical. (2) Else, we have $\algo{PRF}_{k_\epsilon}(M) =0$. Since $\mathbf{F}_\textsc{M}$ is computed identically in both hybrids, and by \Cref{theorem:sampleleft,theorem:sampleright} both $\algo{SampleLeft}$ and $\algo{SampleRight}$ sample from distributions statistically close to $\mathcal{D}_{\Lambda_q^\perp(\mathbf{F}_\textsc{m}),s}$, i.e., they are also statistically close to each other. Thus overall the distributions of signatures returned in $H_3$ and $H_4$ are statistically close to each other, say with less than distance $\Delta(\SecPar)$ (which is negligible in $\SecPar$). Now since \Adv is a quantum machine making at most polynomially (say $q(\secpar)$) many quantum queries. Then, we can use \Cref{lemma:oracleindist} to conclude that \Adv distinguishes between $H_3$ and $H_4$ with probability at most $\sqrt{8C_0q^3\Delta}$, which is negligible. 

\para{Hybrid $H_5$:} In this hybrid, the challenger no longer samples $\mathbf{A}$ using $\algo{TrapGen}$. Instead, it samples $\mathbf{A}$ uniformly from $\mathbb{Z}_q^{n \times m}$. 

\subpara{$H_4 \sind H_5$:} This follows immediately from \Cref{lemma:trapgen}.


\para{Reduction to QSIS.} We can now describe our reduction $\mathcal{R}$ in this hybrid:
\begin{enumerate}
	\item Asks for and recieves a uniform matrix in $\mathbb{Z}_q^{n \times m}$ as the $\mathbf{SIS}_{q,\beta,n,m}$ challenge.  
	\item Sets $\mathbf{A}$ to be this matrix (instead of sampling $\mathbf{A}$ by itself).
	\item When the adversary returns a forgery $(M^*,\sigma^*)$, $\mathcal{R}$ checks if this is valid, i.e., that (i) $M^* \in B_\epsilon$, (ii) $\sigma^* \in \mathbb{Z}_q^{2m}$, (iii) $\sigma^*\neq 0$, (iv) $\mathbf{F}_{\textsc{m}^*} \cdot \sigma^* = 0 \mod q$ and (v) $||\sigma|| \leq \mathsf{sigsize}_\secpar$. If any of these checks fail, it aborts.
	\item Represent $\sigma^*$ as $[\mathbf{d}^\top_1 |~ \mathbf{d}^\top_2]^\top$, with $\mathbf{d}_1, \vb{d}_2 \in \mathbb{Z}_q^m$. $\mathcal{R}$ computes $\mathbf{e} = \mathbf{d}_1 + \mathbf{R}\mathbf{d}_2$ where $\mathbf{R} = \mathbf{R}_{\vb{A}'} - \vb{R}_\textsc{prf,m}$ (we will use this shorthand going forward), and presents $\mathbf{e}$ as its solution to the SIS challenge $\mathbf{A}$.  
\end{enumerate}
Now we can prove that $\vb{e}$ is indeed an SIS solution with non-negligible probability by an  argument very similar as in the final reduction for \cite[Theorem 3.1]{AC:BoyLi16}. We present the final reduction in the following.

\subpara{The Final Reduction.} 
Before showing that the reduction's output $\mathbf{e}$ indeed breaks the given SIS challenge, we must first examine the possibility of a `related message' attack. Namely, we want to avoid a situation where the adversary can directly use signatures for one message to get signatures on another since this would render our reduction moot. We show that this is not the case by showing that an adversary cannot come up with two messages $M,M'$ such that $\mathbf{F}_\textsc{m} = \mathbf{F}_\textsc{m}'$. The following lemma accomplishes this task. 
\begin{lemma}
If a QPT adversary produces two distinct messages $M,M'$ such that $\mathbf{A}_\textsc{prf,m} = \mathbf{A}_{\textsc{prf},\textsc{m}'}$ with non-negligible probability, then we can break the $\mathbf{SIS}_{q,\beta,n,m}$ challenge. 
\end{lemma}  

\begin{proof}
With the verification key in $H_5$ picked just as in $H_2$, if $\mathbf{A}_\textsc{prf,m} = \mathbf{A}_{\textsc{prf},\textsc{m}'}$, then we have $$\mathbf{AR}_\textsc{prf,m} + \algo{PRF}_{k_\epsilon}(M)\mathbf{G} = \mathbf{AR}_{\textsc{prf}, \textsc{m}'} + \algo{PRF}_{k_\epsilon}(M')\mathbf{G}.$$ 
Note that we have $\algo{PRF}_{k_\epsilon}(M) \neq \algo{PRF}_{k_\epsilon}(M')$ with probability $2\epsilon \cdot (1-\epsilon)$, which is a constant. If this holds, we have $\mathbf{A}(\mathbf{R}_\textsc{prf,m} - \mathbf{R}_{\textsc{prf}, \textsc{m}'}) \pm \mathbf{G} = 0 \mod q$. Now by \Cref{lemma:gadgettrapdoor} and using $\algo{SampleRight}$ we can find a low-norm vector $\mathbf{d} \in \mathbb{Z}_q^{m \times m}$ such that $\mathbf{Gd} = 0 \mod q$, $\mathbf{d} \neq 0$ and $||\mathbf{d}|| \leq s'\sqrt{2m}$ (for some $s' \geq \sqrt{5}\omega(\sqrt{\log m})$). Then $[\mathbf{A}(\mathbf{R}_\textsc{prf,m} - \mathbf{R}_{\textsc{prf}, \textsc{m}'}) \pm \mathbf{G}]\mathbf{d} = 0 \mod q$, yielding $\mathbf{A}(\mathbf{R}_\textsc{prf,m} - \mathbf{R}_{\textsc{prf}, \textsc{m}'})\mathbf{d} = 0 \mod q$. By our choice of parameters, $(\mathbf{R}_\textsc{prf,m} - \mathbf{R}_{\textsc{prf}, \textsc{m}'})$ has low enough norm and so $(\mathbf{R}_\textsc{prf,m} - \mathbf{R}_{\textsc{prf}, \textsc{m}'})\mathbf{d}$ is a valid SIS solution for $\mathbf{A}$. This happens with non-negligible probability using our starting assumption, and thus we break $\mathbf{SIS}_{q,\beta,n,m}$ as claimed.
\end{proof}

Now we can turn to validating our reduction. It is straightforward to verify that if $\sigma^*$ is a valid signature, then $\mathbf{e}$ is a valid integer solution to $\mathbf{A}$. Indeed, we have $\mathbf{F}_{\textsc{m}^*}\cdot \sigma^* = 0 \mod q$, which from the above boils down to 
$$[\mathbf{A}~|~\mathbf{A}(\mathbf{R_{A'}}-\mathbf{R}_\textsc{prf,m}) + \big(1-\algo{PRF}_{k_\epsilon}(M)\big)\mathbf{G}] \cdot \sigma^* = 0 \mod q,$$
which can be rewritten as $\mathbf{A}(\mathbf{d}_1 + \mathbf{R}\mathbf{d}_2) = 0 \mod q$, proving our claim. 

It remains to verify that $\mathbf{e}$ is (i) short and (ii) nonzero. Let us begin with shortness. Since $||\sigma^*|| \leq s\sqrt{2m}$, we have $\mathbf{||d_1||,||d_2||} \leq s\sqrt{2m}$. We then have $||\mathbf{e}|| \leq ||\mathbf{d}_1|| + ||\mathbf{d}_2||\cdot||\mathbf{R}||_2$. By our parameter choices, and using \Cref{lemma:rnorm}, this latter term is again at most $O(4^d m^{3/2})s\sqrt{2m}$. By our choice of parameters, this is less that $\beta \ge O(4^d\cdot m^{3/2}) \cdot s\sqrt{m}$, and so $\mathbf{e}$ is indeed a valid solution. 

Next let us show that $\mathbf{e}$ is nonzero with overwhelming probability. Note that by assumption, $\sigma^*$ is nonzero so at least one of $\mathbf{d}_1$ or $\mathbf{d}_2$ must be so. If $\mathbf{d}_2$ is zero, then we have that $\mathbf{e}$ is directly is nonzero, so let us focus on the case that $\mathbf{d}_2$ is nonzero. Expressing $\mathbf{d}_2$ as $(d_1,\dots,d_m)^\top$, we must have that at least one of the coordinates of $\mathbf{d}_2$ is nonzero. Let $d_j$ be such a coordinate. Expressing $\mathbf{R}$ as $(\mathbf{r}_1,\dots,\mathbf{r}_m)$, we have that $$\mathbf{R}\cdot\mathbf{d}_2 = \mathbf{r}_jd_j + \sum\limits_{i=1,i\neq j}^{m}\mathbf{r}_id_i.$$ 
Now we note that for the (fixed) $M^*$ for which $\Adv$ makes its forgery, $\mathbf{R}$ (and in turn $\mathbf{r}_j$) depends only on the low-norm matrices $\mathbf{R_{A'}},\Set{\mathbf{R_{B_i}}}_{i \in [k]},\mathbf{R_{C_0}},\mathbf{R_{C_1}}$ and $k_\epsilon$. Now the {\em only} information about $\mathbf{R}$ (in turn, $\mathbf{r}_j$) $\Adv$ has is derived from the components of $vk$, namely, $\mathbf{A}',\Set{\mathbf{B}_i}_{i \in [k]},\mathbf{C}_0,\mathbf{C}_1$. This implies that {\em any} $\mathbf{r}'_j \in \Set{-1,1}^m$ such that $\mathbf{Ar}_j = \mathbf{Ar}'_j$ is in fact a valid vector in the sense that replacing $\mathbf{r}_j$ with $\mathbf{r}'_j$ is completely indistinguishable to the adversary. By the pigeonhole principle, there are exponentially many such distinct $\mathbf{r}'_j$'s so that $\mathbf{Ar}_j = \mathbf{Ar}'_j$, and for such an admissible $\mathbf{r}'_j$, the probability that $\mathbf{r}'_j\cdot d_j$ hits a fixed value in $\mathbb{Z}^m_q$ is exponentially small. It is straightforward to see that this implies that $\mathbf{e}$ is zero with at most negligible probability (since the chance that $\mathbf{r}'_j\cdot d_j$ equals the exact value needed to cancel out the other terms in $\mathbf{e}$ is negligible). 
Finally, it is straightforward to verify that $H_4$ runs in polynomial time, and in turn that the reduction $\mathcal{R}$ is efficient. If $\Adv$ produces a valid forgery within $B_\epsilon$ with probability $\nu(\secpar)$, $\mathcal{R}$ breaks the $\mathbf{SIS}_{q,\beta,n,m}$ challenge with probability $\nu(\secpar)-\negl(\secpar)$. We conclude that $\Adv$ wins the blind-unforgeability experiment with at most negligible probability. 
\end{proof}


\section{Post-Quantum Ring Signatures}
\subsection{Definition}\label{sec:PQ-ring-sig:def}
\subsubsection{Classical Ring Signatures} 
We start by recalling the classical definition of ring signatures \cite{TCC:BenKatMor06,EC:BDHKS19}.

\begin{definition}[Ring Signature]
	\label{def:classical:ring-signature}
	A ring signature scheme \algo{RS} is described by a triple of PPT algorithms \algo{(Gen,Sign,Verify)} such that:
	\begin{itemize}
		
		\item {\bf $\Gen(1^\lambda,N)$:} on input a security parameter $1^\lambda$ and a super-polynomial\footnote{The $N$ has to be super-polynomial to support rings of {\em arbitrary} polynomial size.} $N$ (e.g., $N = 2^{\log^2\secpar}$) specifying the maximum number of members in a ring, output a verification and signing key pair $\algo{(VK,SK)}$. 
		
		\item {\bf $\algo{Sign}(\SK,\Ring, m)$:} given a secret key \SK, a message $m \in \mathcal{M}_\lambda$, and a list of verification keys (interpreted as a ring) $\algo{R = (VK_1,\cdots,VK_\ell)}$ as input, and outputs a signature $\Sigma$. 
		
		\item {\bf $\algo{Verify}(\Ring,m,\Sigma)$:} given a ring $\algo{R = (VK_1,\dots,VK_\ell)}$, message $m \in \mathcal{M}_\lambda$ and a signature $\Sigma$ as input, outputs either 0 (rejecting) or 1 (accepting).  
	\end{itemize}
	These algorithms satisfy the following requirements:  
	\begin{enumerate}
		\item {\bf Completeness:} for all $\lambda \in \Naturals$, $\ell \le N$, $i^* \in [\ell]$, and $m \in \mathcal{M}_\lambda$, it holds that $\forall i \in [\ell]$ $(\VK_i,\SK_i) \gets \Gen(1^\lambda,N)$ and $\Sigma \gets \algo{Sign}(\SK_{i^*},\Ring,m)$ where $\algo{R = (VK_1,\dots,VK_\ell)}$, we have 
		$$\Pr[\algo{RS.Verify}(\Ring,m,\Sigma) = 1] = 1,$$ 
		where the probability is taken over the random coins used by $\Gen$ and $\algo{Sign}$.
		
		\item {\bf Anonymity:} For any $Q = \poly(\secpar)$ and any PPT adversary $\Adv$, it holds w.r.t.\ \Cref{expr:classical-anonymity} that 
		$$\algo{Adv}_\textsc{Anon}^{\secpar, Q}(\Adv) \coloneqq \big|\Pr\big[\algo{Exp}_\textsc{Anon}^{\secpar, Q}(\Adv) = 1\big] - 1/2\big| \le \negl(\secpar).$$
		\begin{ExperimentBox}[label={expr:classical-anonymity}]{Classical Anonymity \textnormal{$\algo{Exp}_\textsc{Anon}^{\secpar, Q}(\Adv)$}}
		\begin{enumerate}[label={\arabic*.},leftmargin=*,itemsep=0em]
			\item 
			For each $i \in [Q]$, the challenger generates key pairs $(\VK_i,\SK_i) \gets \Gen(1^\lambda, N;r_i)$. It sends $\Set{(\VK_i, \SK_i, r_i)}_{i\in [Q]}$ to $\Adv$;
			\item 
			$\Adv$ sends a challenge to the challenger of the form $(i_0,i_1,\Ring,m)$.\footnote{We stress that $\Ring$ might contain keys that are not generated by the challenger in the previous step. In particular, it might contain maliciously generated keys.} The challenger checks if $\VK_{i_0} \in \Ring$ and $\VK_{i_1} \in \Ring$. If so, it samples a uniform bit $b$, computes $\Sigma \gets \algo{Sign}(\SK_{i_b},\Ring,m)$, and sends $\Sigma$ to $\Adv$. 
			\item 
			$\Adv$ outputs a guess $b'$. If $b' = b$, the experiment outputs 1, otherwise 0.
		\end{enumerate}
		\end{ExperimentBox}

		\item {\bf Unforgeability:} for any $Q=\poly(\lambda)$ and any PPT adversary $\Adv$, it holds w.r.t.\ \Cref{expr:classical:unforgeability} that 
		$$\algo{Adv}_\textsc{Unf}^{\secpar,  Q}(\Adv) \coloneqq \Pr\big[\algo{Exp}_\textsc{Unf}^{\secpar,  Q}(\Adv) = 1\big] \le \negl(\secpar).$$

		\begin{ExperimentBox}[label={expr:classical:unforgeability}]{Classical Unforgeability \textnormal{$\algo{Exp}_\textsc{Unf}^{\secpar, Q}(\Adv)$}}
		\begin{enumerate}[label={\arabic*.},leftmargin=*,itemsep=0em]
			\item For each $i \in [Q]$, the challenger generates $(\VK_i,\SK_i) \gets \Gen(1^\lambda,N;r_i)$, and stores these key pairs along with their corresponding randomness. It then sets $\mathcal{VK} = \Set{\VK_1,\dots,\VK_Q}$ and initializes a set $\mathcal{C} = \emptyset$. 
			\item The challenger sends $\mathcal{VK}$ to $\Adv$. 
			\item $\Adv$ can make polynomially-many queries of the following two types: 
			\begin{itemize}[leftmargin=*,itemsep=0em,topsep=0em]
			\item {\bf Corruption query $(\mathsf{corrupt},i)$:} The challenger adds $\VK_i$ to the set $\mathcal{C}$ and returns the randomness $r_i$ to \Adv. 
			\item {\bf Signing query $(\algo{sign},i,\Ring,m)$:} The challenger first checks if $\VK_i \in \Ring$. If so, it computes $\Sigma \gets \algo{Sign}(\SK_i,\Ring,m)$ and returns $\Sigma$ to \Adv. It also keeps a list of all such queries made by \Adv. 
			\end{itemize}   
			\item Finally, $\Adv$ outputs a tuple $(\Ring^*, m^*, \Sigma^*)$. The challenger checks if: 
			\begin{enumerate}
			\item
			 $\Ring^* \subseteq \mathcal{VK \setminus C}$,
			\item
			 $\Adv$ never made a signing query of the form $(\algo{sign},\cdot,\Ring^*, m^*)$, {\bf and} 
			 \item
			  $\algo{Verify}(\Ring^*,m^*,\Sigma^*) = 1$.
			 \end{enumerate}
			If so, the experiment outputs 1; otherwise, 0.
		\end{enumerate}
		\end{ExperimentBox}		
	\end{enumerate}
\end{definition}
We mention that the unforgeability and anonymity properties defined in \autoref{def:classical:ring-signature} correspond respectively to the notions of \emph{unforgeability with insider corruption} and \emph{anonymity with respect to full key exposure} presented in \cite{TCC:BenKatMor06}.

\subsubsection{Defining Post-Quantum Security}\label{sec:ring-sig:pq-definition}
We aim to build a classical ring signature that is secure against adversaries making superposition queries to the signing oracle. Formalizing the security requirements in this scenario is non-trivial. An initial step toward this direction has been taken in \cite{C:CGHKLMPS21}. But their definition has certain restrictions (discussed below). In the following, we develop a new definition building on ideas from \cite{C:CGHKLMPS21}.

\para{Post-Quantum Anonymity.} Recall that in the classical anonymity game (\Cref{expr:classical-anonymity}), the adversary's challenge is a quadruple $(i_0,i_1,\Ring,m)$. To define post-quantum anonymity, a natural attempt is to allow the adversary to send a superposition over components of quadruple, and to let the challenger respond using the following unitary mapping\footnote{Of course, the challenger also needs to check if $\VK_{i_0} \in \Ring$ and $\VK_{i_1} \in \Ring$. But we can safely ignore this for our current discussion.}:
$$\sum_{i_0,i_1,\Ring,m,t} \psi_{i_0, i_1, \Ring, m,  t}\ket{i_0, i_1, \Ring, m, t} \mapsto \sum_{i_0,i_1,\Ring,m,t} \psi_{i_0, i_1, \Ring, m, t}\ket{i_0, i_1, \Ring, m, t \oplus \Sign(\SK_{i_b}, m, \Ring;r)}.$$ 
However, as observed in \cite{C:CGHKLMPS21}, this will lead to an unsatisfiable definition due to an attack from \cite{C:BonZha13}. Roughly speaking, the adversary could use classical values for $\Ring$, $m$, and $i_1$, but she puts a uniform superposition of all valid identities in the register for $i_0$. After the challenger's signing operation, observe that if $b = 0$, the last register will contain  signatures in superposition (as $i_0$ is in superposition); if $b = 1$, it will contain a classical signature (as $i_1$ is classical). These two cases can be efficiently distinguished by means of a Fourier transform on the $i_0$'s register followed by a measurement. Therefore, to obtain an achievable notion, we should not allow superpositions over $(i_0, i_1)$. 

Now, $\Adv$ only has the choice to put superpositions over $\Ring$ and $m$. The definition in \cite{C:CGHKLMPS21} further forbids $\Adv$ from putting superpositions over $\Ring$. But this is only because they fail to prove security if superposition attacks on $\Ring$ is allowed. Indeed, they leave open the problem to construct a scheme that protects against superposition attacks on $\Ring$. In this work, we solve this problem: our definition allows superposition attacks on both $\Ring$ and $m$.
\begin{definition}[Post-Quantum Anonymity]
\label{def:pq:anonymity}
Consider a triple of PPT algorithms $\RS = (\algo{Gen}, \Sign, \Verify)$ that satisfies the same syntax as in \Cref{def:classical:ring-signature}. $\RS$ achieves post-quantum anonymity if for any $Q=\poly(\lambda)$ and any QPT adversary $\Adv$, it holds w.r.t.\ \Cref{expr:pq:anonymity} that 
		$$\algo{PQAdv}_\textsc{Anon}^{\secpar,Q}(\Adv) \coloneqq \big|\Pr\big[\algo{PQExp}_\textsc{Anon}^{\secpar,Q}(\Adv) = 1\big] - 1/2\big| \le \negl(\secpar).$$
\begin{ExperimentBox}[label={expr:pq:anonymity}]{Post-Quantum Anonymity \textnormal{$\algo{PQExp}_\textsc{Anon}^{\secpar,Q}(\Adv)$}}
\begin{enumerate}
\item 
For each $i \in [Q]$, the challenger generates key pairs $(\VK_i,\SK_i) \leftarrow \RS.\algo{Gen}(1^\SecPar, N;r_i)$. The challenger sends $\Set{(\VK_i, \SK_i, r_i)}_{i\in [Q]}$ to $\Adv$;
\item
$\Adv$ sends $(i_0, i_1)$ to the challenger, where both $i_0$ and $i_1$ are in $[Q]$;
\item
$\Adv$'s challenge query is allowed to be a superposition of rings {\em and} messages. The challenger picks a random bit $b$ and a random string $r$. It signs the message using $\SK_{i_b}$ and randomness $r$, while making sure that $\VK_{i_0}$ and $\VK_{i_1}$ are indeed in the ring specified by $\Adv$. Formally, the challenger implements the following mapping:
$$\sum_{\Ring, m, t}\psi_{\Ring, m, t} \ket{ \Ring, m, t} \mapsto \sum_{\Ring, m, t}\psi_{\Ring, m, t} \ket{\Ring, m, t\xor f(\Ring, m)},$$
where
$f(\Ring, m) \coloneqq
\begin{cases}
\RS.\Sign(\SK_{i_b}, R, m;r) & \text{if}~\VK_{i_0}, \VK_{i_1} \in R \\
\bot & \text{otherwise}
\end{cases}
$.
\item 
$\Adv$ outputs a guess $b'$. If $b' = b$, the experiment outputs 1, otherwise 0.
\end{enumerate}
\end{ExperimentBox}
\end{definition}

\para{Post-Quantum Unforgeability.} In the classical unforgeability game (\Cref{expr:classical:unforgeability}), $\Adv$ can make both {\sf corrupt} and {\sf sign} queries. As discussed in \Cref{sec:tech-overview:PQ-ring-sig}, we do not consider quantum {\sf corrupt} queries, or superposition attacks over the identity in $\Adv$'s {\sf sign} queries. 
We also remark that in the unforgeability game, \cite{C:CGHKLMPS21} does not allow superpositions over the ring. Instead of a definitional issue, this is again only because they are unable to prove the security of their scheme if superposition attacks on the ring is allowed. In contrast, our construction can be proven secure against such attacks; thus, this restriction is removed from our definition.



To define quantum unforgeability, \cite{C:CGHKLMPS21} adapts one-more unforgeability \cite{C:BonZha13} to the ring setting: they require that, with $\mathsf{sq}$ quantum signing queries, the adversary cannot produce $\mathsf{sq} + 1$ signatures, where all the rings are subsets of $\mathcal{VK}\setminus \mathcal{C}$. This definition, {\em when restricted to the classical setting}, seems to be weaker than the standard unforgeability in \Cref{def:classical:ring-signature}.
 That is, in the classical setting, any $\RS$ satisfying the unforgeability in \Cref{def:classical:ring-signature} is also one-more unforgeable; but the reverse direction is unclear (we discuss this in \Cref{sec:one-more:PQ-EUF:ring-sig}). Instead, our definition extends the blind-unforgeability for ordinary signatures (\Cref{def:classical-BU-sig}) to the ring setting. We present this definition in \Cref{def:pq:blind-unforgeability}. 
\begin{definition}[Post-Quantum Blind-Unforgeability]
\label{def:pq:blind-unforgeability}
Consider a triple of PPT algorithms $\RS = (\algo{Gen}, \Sign, \Verify)$ that satisfies the same syntax as in \Cref{def:classical:ring-signature}. For any security parameter $\secpar$, let $\mathcal{R}_\secpar$ and $\mathcal{M}_\secpar$ denote the ring space and message space, respectively. $\RS$ achieves blind-unforgeability if for any $Q=\poly(\lambda)$ and any QPT adversary $\Adv$, it holds w.r.t.\ \Cref{expr:pq:blind-unforgeability} that 
		$$\algo{PQAdv}_\textsc{bu}^{\secpar,Q}(\Adv) \coloneqq \Pr\big[\algo{PQExp}_\textsc{bu}^{\secpar,Q}(\Adv) = 1\big] \le \negl(\secpar).$$ 
\begin{ExperimentBox}[label={expr:pq:blind-unforgeability}]{Post-Quantum Blind-Unforgeability \textnormal{$\algo{PQExp}_\textsc{bu}^{\secpar,Q}(\Adv)$}}
\begin{enumerate}
	\item 
	$\Adv$ sends a constant $0\le \epsilon \le 1$ to the challenger;
			
	\item 
	For each $i \in [Q]$, the challenger generates $(\VK_i,\SK_i) \gets \Gen(1^\lambda,N;r_i)$, and stores these key pairs along with their corresponding randomness. It then sets $\mathcal{VK} = \Set{\VK_1,\dots,\VK_Q}$ and initializes a set $\mathcal{C} = \emptyset$; The challenger sends $\mathcal{VK}$ to $\Adv$; 

	\item 
	The challenger defines a {\em blindset} $B^\RS_\varepsilon \subseteq 2^{\mathcal{R}_\secpar} \times \mathcal{M}_\SecPar$: every pair $(\Ring, m) \in 2^{\mathcal{R}_\secpar} \times \mathcal{M}_\SecPar$ is put in $B^\RS_\varepsilon$ with probability $\varepsilon$; 
	
	\item 
	$\Adv$ can make polynomially-many queries of the following two types:  
	\begin{itemize}
	\item 
	{\bf Classical corruption query $(\mathsf{corrupt},i)$:} The challenger adds $\VK_i$ to the set $\mathcal{C}$ and returns the randomness $r_i$ to \Adv. 
	\item 
	{\bf Quantum Signing query $(\algo{sign},i, \sum \psi_{\Ring, m, t}\ket{\Ring, m, t})$:} That is, $\Adv$ is allowed to query the signing oracle on some classical identity $i$ and  superpositions over rings and messages. The challenger samples a random string $r$ and performs:
	$$\sum_{\Ring, m, t}\psi_{\Ring, m, t} \ket{\Ring, m, t} \mapsto \sum_{\Ring, m, t}\psi_{\Ring, m, t} \ket{\Ring, m, t\xor B^\RS_\epsilon f(\Ring, m)},$$
	where $B^\RS_\epsilon f(\Ring, m) \coloneqq
	\begin{cases}
	 \bot  & \text{if}~(\Ring, m) \in B^\RS_\epsilon \\
	f(\Ring, m) & \text{otherwise}
	\end{cases}
	$, and

	$f(\Ring, m) \coloneqq
	\begin{cases}
	\RS.\Sign(\SK_{i}, m, \Ring;r) & \text{if}~\VK_{i} \in \Ring \\
	\bot & \text{otherwise}
	\end{cases}
	$.
	\end{itemize}   
	\item Finally, $\Adv$ outputs $(\Ring^*, m^*, \Sigma^*)$. The challenger checks if: 
	\begin{enumerate}
	\item
	 $\Ring^* \subseteq \mathcal{VK \setminus C}$, 
	\item
	 $\algo{Verify}(\Ring^*,m^*,\Sigma^*) = 1$, {\bf and}
	\item
	$(\Ring^*,m^*) \in B^\RS_\epsilon$.
	\end{enumerate}
	If so, it outputs 1; otherwise, it outputs 0.
\end{enumerate}
\end{ExperimentBox}
\end{definition}

In contrast to the ``one-more'' unforgeability, we show in \Cref{lem:euf-ring-sig:bu-ring-sig} that, when restricted to the classical setting, \Cref{def:pq:blind-unforgeability} (for ring signatures) is indeed equivalent to the standard existential unforgeability in \Cref{def:classical:ring-signature}. Its proof is almost identical to that of \cite[Proposition 2]{EC:AMRS20}. 
\begin{lemma}\label{lem:euf-ring-sig:bu-ring-sig}
Restricted to (classical) QPT adversaries, a ring signature $\RS$ scheme is blind-unforgeable (\Cref{def:pq:blind-unforgeability}) if and only if it satisfies the unforgeability requirement in \Cref{def:classical:ring-signature}.
\end{lemma}
\begin{proof}

We show necessity and sufficiency in turn. In the following, by ``\Cref{expr:pq:blind-unforgeability}'', we refer to the classical version of \Cref{expr:pq:blind-unforgeability}, where the signing query is of the form $(\algo{sign}, i, \Ring, m)$ (i.e., $(\Ring, m)$ is classical), and is answered as $B^\RS_\epsilon f(\Ring, m)$.	

\subpara{Necessity ($\Leftarrow$).} Let us first show how blind-unforgeability implies  standard unforgeability (for classical settings). Assume we have an adversary $\Adv_\textsc{euf}$ that breaks standard unforgeability of $\RS$ as per  \Cref{def:classical:ring-signature}, i.e., in \Cref{expr:classical:unforgeability} it produces a forgery $(m^*,\Ring^*,\Sigma^*)$   that is valid with non-negligible probability $\nu(\SecPar)$. We show that this is easily converted into an adversary $\Adv_\textsc{bu}$ that wins \Cref{expr:pq:blind-unforgeability} with non-negligible probability as well. $\Adv_\textsc{bu}$ first sets $\epsilon(\SecPar)$ equal to $1/p(\SecPar)$, where $p(\secpar)$ denotes the (polynomial) running time of $\Adv_\textsc{euf}$ (the reasoning behind this choice will become clear very soon). It then simply forwards all the queries from $\Adv_\textsc{euf}$ to the blind-unforgeability challenger and the responses back to $\Adv_\textsc{euf}$. It also outputs whatever eventual forgery $\Adv_\textsc{euf}$ does. 

Let us consider the success probability of $\Adv_\textsc{bu}$. To start with, note that $\Adv_\textsc{euf}$ makes at most $p(\SecPar)$ many queries of the signing oracle. In each such query, we know that the 
$(\Ring, m)$ pair is in the blind set independently with probability $\epsilon$. Thus it is not in the blind set with probability $1-\epsilon$, and if so the query is answered properly. In turn, the probability that all the queries made are answered properly is then at least $(1-\epsilon)^{p(\SecPar)} \approx 1/e$ (this uses independence and $\epsilon = 1/p$), and so the probability that the forgery $(\Ring^*,m^*,\Sigma^*)$ is valid 
is then at least $(1-\epsilon)^{p(\SecPar)}\cdot\nu(\SecPar)$.
Finally, the forgery, even if successful, might lie in the blind set with probability $\epsilon$. So, the total probability that $\Adv_\textsc{euf}$ outputs a valid forgery {\em for the blind unforgeability game} is $(1-\epsilon)^{p(\SecPar)+1} \cdot \nu(\SecPar) \approx (1-\epsilon) \cdot \nu \cdot 1/e$, which is non-negligible since $\nu$ is non-negligible by assumption. Thus if $\Adv_\textsc{euf}$ violates standard ring signature unforgeability according to \Cref{def:classical:ring-signature}, then $\Adv_\textsc{bu}$ violates blind unforgeability for ring signatures according to \Cref{def:pq:blind-unforgeability}, as claimed. 

\subpara{Sufficiency ($\Rightarrow$).} Let us now turn to the other direction of the equivalence. Assume now that there exists an adversary $\Adv_\textsc{bu}$ that can break blind unforgeability of $\RS$, i.e., win \Cref{expr:pq:blind-unforgeability} with non-negligible probability $\nu(\secpar)$. We show an adversary $\Adv_\textsc{euf}$
that can win \Cref{expr:classical:unforgeability} with non-negligible probability. $\Adv_\textsc{euf}$ simply simulates \Cref{expr:pq:blind-unforgeability} for $\Adv_\textsc{bu}$ by answering oracle queries according to a locally-simulated version of $B^\RS_\epsilon f(\Ring,m)$. Concretely, the adversary $\Adv_\textsc{euf}$ proceeds by drawing a subset $B^\RS_\epsilon$ in the same manner as the challenger in \Cref{expr:pq:blind-unforgeability}  and answering queries made by $\Adv_\textsc{bu}$ according to $B^\RS_\epsilon f(\Ring,m)$. Two remarks are in order:
\begin{enumerate}
\item \label{item:euf-ring-sig:bu-ring-sig:remark1}
when $(\Ring, m) \in B^\RS_\secpar$, no signature needs to be done. That is, this query can be answered by $\Adv_\textsc{euf}$ without calling its own signing oracle;
\item
$\Adv_\textsc{euf}$ can construct the set $B^\RS_\epsilon$ by ``lazy sampling'', i.e., when a particular query $(\algo{sign}, i, \Ring, m)$ is made by $\Adv_\textsc{bu}$, 
whether $(\Ring, m) \in B^\RS_\epsilon$ and ``remembering'' this information in case the query is asked again.
\end{enumerate}
By assumption, $\Adv_\textsc{bu}$ produces a valid forgery. And it follows from \Cref{item:euf-ring-sig:bu-ring-sig:remark1} that this forgery must be on a point which was not queried by $\Adv_\textsc{euf}$, thus, also serving as a valid forgery for $\Adv_\textsc{euf}$'s game.
\end{proof}

\noindent{To conclude, we present the complete definition for quantum ring signatures.}
\begin{definition}[Post-Quantum Secure Ring Signatures]
\label{def:pq:ring-signatures}
A post-quantum secure ring signature scheme $\RS$ is described by a triple of PPT algorithms $(\Gen,\Sign,\Verify)$ that share the same syntax as in \Cref{def:classical:ring-signature}. Moreover, these algorithms also satisfy the {\em completeness} requirement as per \Cref{def:classical:ring-signature}, the {\em post-quantum anonymity} requirement as per \Cref{def:pq:anonymity}, and the {\em post-quantum blind-unforgeability} requirement as per \Cref{def:pq:blind-unforgeability}.
\end{definition}

\subsection{Building Blocks}

\subsubsection{Lossy PKEs with Special Properties} 
We need the following lossy PKE.
\begin{definition}[Special Lossy PKE]\label{def:special-LE}
For any security parameter $\secpar \in \Naturals$, let $\mathcal{M}_\secpar$ denote the message space.
A special lossy public-key encryption scheme $\algo{LE}$ consists of the following PPT algorithms:
\begin{itemize}

\item
$\algo{MSKGen}(1^\secpar, Q)$, on input a number $Q \in \Naturals$, outputs $\big(\Set{\pk_i}_{i \in [Q]}, \msk \big)$. We call $\pk_i$'s the {\em injective public keys}, and $\msk$ the {\em master secret key}.
\item
$\algo{MSKExt}(\msk, \pk)$, on input a master secret key $\msk$ and an injective public key $\pk$, outputs a secret key $\sk$. 

\item
 $\KSam^\ls(1^\secpar)$ outputs key $\pk_\ls$, which we call {\em lossy public key}.
     
\item 
$\algo{Valid}(\pk,\sk)$, on input a public $\pk$ and a secret key $\sk$, outputs either $1$ (accepting) or $0$ (rejecting).

\item
$\RndExt(\pk)$ outputs a $r$ which we call {\em extracted randomness}.

\item 
$\algo{Enc}(\pk, m)$, on input a public key $\pk$, and a message $m \in \mathcal{M}_\secpar$, outputs $\ct$.

\item 
$\algo{Dec}(\sk, \ct)$, on input a secret key $\sk$ and a ciphertext $ct$, outputs $m$.
\end{itemize}
These algorithms satisfy the following properties:
\begin{enumerate}
\item \label{item:def:LE:completeness}
{\bf Completeness.} For any $\secpar\in\Naturals$, any $(\pk, \sk)$ s.t.\ $\Valid(\pk, \sk) =1$, and any $m \in \mathcal{M}_\secpar$, it holds that 
$$\Pr[\Dec\big(\sk, \Enc(\pk, m)\big)= m] = 1.$$

\item \label{def:LE:property:lossyness}
{\bf Lossiness of lossy keys.} For any $\pk_\ls$ in the range of $\KSam^\ls(1^\secpar)$ and any $m_0, m_1 \in \mathcal{M}_\secpar$, it holds that
$$\big\{ \Enc(\pk_\ls, m_0) \big\}_{\secpar\in\Naturals} \sind \big\{ \Enc(\pk_\ls, m_1) \big\}_{\secpar\in\Naturals}.$$



\item \label{def:LE:property:GenWithSK:completeness}  
{\bf Completeness of Master Secret Keys:} for any $Q = \poly(\secpar)$, it holds that
$$\Pr[
\big(\Set{\pk_i}_{i \in [Q]}, \algo{msk} \big) \gets \algo{MSKGen}(1^\secpar, Q): 
\begin{array}{l}
\forall i \in [Q], \Valid(\pk_i,\sk_i\big) =1, \\
\textrm{where}~\sk_i \coloneqq \algo{MSKExt}\big(\algo{msk}, \pk_i)
\end{array}
] \ge 1 - \negl(\secpar).$$

\item \label{def:LE:property:IND:GenWithSK-KSam}  
{\bf IND of $\algo{MSKGen}$/$\KSam^\ls$ mode:} For any $Q=\poly(\secpar)$, the following two distributions are computationally indistinguishable:
\begin{itemize}
\item
$\forall i\in [Q]$, sample $\pk_i \gets \KSam^\ls(1^\secpar; r_i)$, then output $\Set{\pk_i, r_i}_{i\in [Q]}$;
\item 
Sample $\big(\Set{\pk_i}_{i \in [Q]}, \algo{msk} \big) \gets \algo{MSKGen}(1^\secpar, Q)$ and output $\big\{\pk_i, \algo{RndExt}(\pk_i)\big\}_{i\in [Q]}$.
\end{itemize}

\item \label{item:def:LE:computationally-unique-sk}
{\bf Almost-Unique Secret Key:} For any $Q = \poly(\secpar)$, it holds that
$$
\Pr[
\big(\Set{\pk_i}_{i \in [Q]}, \algo{msk} \big) \gets \algo{MSKGen}(1^\secpar, Q): 
\begin{array}{l}
\text{There exist}~ i \in [Q] ~\text{and}~ \sk'_i~\text{such that} \\
\sk'_i \ne \algo{MSKExt}(\algo{msk},\pk_i) ~\wedge~ \Valid(\pk_i, \sk'_i) = 1
 \end{array}
] = \negl(\secpar).$$
\end{enumerate}
\end{definition}
We propose an instantiation of such a lossy PKE using dual mode LWE commitments \cite{STOC:GorVaiWic15}. In lossy (statistically hiding) mode, the public key consists of a uniformly sampled matrix $\mathbf{A}$ and a message $m$ is encrypted by computing $\mathbf{A}\mathbf{R} + m\mathbf{G}$, where $\mathbf{R}$ is a low-norm matrix and $\mathbf{G}$ is the gadget matrix. Note that the random coins used to sample $\mathbf{A}$ simply consists of the matrix $\mathbf{A}$ itself. Furthermore, we can switch $\mathbf{A}$ to be an LWE-matrix (using some secret vector $\mathbf{s}$) to make the encryption scheme injective. Such a modification is computationally indistinguishable by an invocation of the LWE assumption. Note that this is true also in the presence of the output of $\algo{RndExt}(\mathbf{A})$, since the algorithm simply returns $\mathbf{A}$. Furthermore, by setting the dimensions appropriately, the secret $\mathbf{s}$ is uniquely determined by $\mathbf{A}$ with overwhelming probability. Finally, we note that we can define a master secret key for all keys in injective mode using a simple trick: sample a PRF key $k$ and sample the $i$-th key pair using $\mathsf{PRF}(k,i)$ as the random coins. It is not hard to see that the distribution of public/secret keys is computationally indistinguishable by the pseudorandomness of $\mathsf{PRF}$. Furthermore, given $k$ one can extract the $i$-th secret key simply by recomputing it.

\subsubsection{ZAPs for Super-Complement Languages}
As mentioned in \Cref{sec:tech-overview:PQ-ring-sig}, \cite{C:CGHKLMPS21} uses a ZAP (for $\NP\cap\coNP$) to prove a statement that the (ring) signature contains a ciphertext of a valid signature w.r.t.\ the building-block signature scheme. Let us denote this language as $L$. In the security proof, they need to argue that the adversary cannot prove a false statement $x^*\notin L$. However, this $L$ is not necessarily in $\coNP$; thus, there may not exist a non-witness $\tilde{w}$ for the fact that $x^*\notin L$. Therefore, it is unclear how to use a ZAP for $\NP\cap \coNP$ here. To address this issue, the authors of \cite{C:CGHKLMPS21} propose the notion of {\em super-complement languages}. This notion considers a pair of $\NP$ languages $(L, \tilde{L})$ such that $(x\in \tilde{L}) \Rightarrow (x \notin L)$. Their ZAP achieves soundness such that the cheating prover cannot prove $x \in L$ (except with negligible probability) once there exists a ``non-witness'' $\tilde{w}$ s.t.\ $(x,\tilde{w}) \in R_{\tilde{L}}$. The $\tilde{L}$ is set to a special language that captures some {\em necessary conditions} for any forged signatures to be valid. Thus, a winning adversary will break the soundness of the ZAP, leading to a contradiction.

In the following, we present the original definition of super-complement languages. But we will only need a special case of it (see \Cref{rmk:super-complement-language:special-case}).
\begin{definition}[Super-Complement \cite{C:CGHKLMPS21}]\label{def:super-complement-language}
    Let $(L, \widetilde{L})$ be two $\NP$ languages where the elements of $\widetilde{L}$ are represented as pairs of bit strings. We say $\widetilde{L}$ is a \emph{super-complement} of $L$, if 
    $\widetilde{L} \subseteq (\bits^*\setminus L) \times \bits^*$. 
    I.e., $\widetilde{L}$ is a super complement of $L$ if for any $x= (x_1,x_2)$, $x \in \widetilde{L} \Rightarrow x_1 \not \in L$.   
\end{definition}
Notice that, while the complement of $L$ might not be in $\NP$, it must hold that $\widetilde{L} \in \NP$. The language $\widetilde{L}$ is used to define the soundness property. Namely, producing a proof for a statement $x = (x_1, x_2) \in \widetilde{L}$, should be hard. We also use the fact that $\widetilde{L} \in \NP$ to mildly strengthen the soundness property. In more detail, instead of having selective soundness where the statement $x \in \widetilde{L}$ is fixed in advance, we now fix a non-witness $\widetilde{w}$ and let the statement $x$ be adaptively chosen by the malicious prover from all statements which have $\widetilde{w}$ as a witness to their membership in $\widetilde{L}$. 

\begin{remark}\label{rmk:super-complement-language:special-case}
Our application only needs a special case of the general form given in \Cref{def:super-complement-language}---we will only focus on $\tilde{L}$ where the $x_2$ part is an empty string. Formally, we consider the special case where $\tilde{L} \subseteq \bits^* \setminus L $ (i.e., $x \in \tilde{L} \Rightarrow x \notin L$). 
\end{remark}

We now define ZAPs for super-complement languages. We remark that the original definition (and construction) in \cite{C:CGHKLMPS21} captures the general $(L, \tilde{L})$ pairs defined in \Cref{def:super-complement-language}. Since we only need the special case in \Cref{rmk:super-complement-language:special-case}, we will define the ZAP only for this case.

\begin{definition}[ZAPs for Special Super-Complement Languages]
    \label{def:generalizedzap-interface}
    Let $L,\widetilde{L} \in \NP$ be the special super-complement language in \Cref{rmk:super-complement-language:special-case}. Let $R$ and $\widetilde{R}$ denote the $\NP$ relations corresponding to $L$ and $\widetilde{L}$ respectively. Let $\{C_{n,\ell}\}_{n,\ell}$ and $\{\widetilde{C}_{{n},\tilde{\ell}}\}_{{n},\tilde{\ell} }$ be the $\NP$ verification circuits for $L$ and $\widetilde{L}$ respectively. Let $\widetilde{d} = \widetilde{d}({n},\tilde{\ell})$  be the depth of $\widetilde{C}_{n,\tilde{\ell}}$. A {\em ZAP} for $(L,\widetilde{L})$
    is a tuple of PPT algorithms
    $(\algo{V}, \algo{P}, \algo{Verify})$ having the following
    interfaces (where $1^{n}, 1^{\lambda}$ are implicit inputs to
    $\algo{P}$, $\algo{Verify}$):
    \begin{itemize}
        \item 
        {\bf $\algo{V}(1^{\lambda}, 1^n,1^{\widetilde{\ell}}, 1^{\widetilde{D}})$:} 
        On input a security parameter $\lambda$, 
        statement length~$n$ for $L$, 
        witness length $\widetilde{\ell}$ for $\widetilde{L}$, 
        and $\NP$ verifier circuit depth upper-bound $\widetilde{D}$ for $\widetilde{L}$, output a first message
        $\rho$.
        \item 
        {\bf $\algo{P}\big(\rho, x, w\big)$:} On input a string~$\rho$, a statement $x \in \bits^{n}$, and a witness $w$ such that $(x, w) \in R$, output a proof $\pi$.
        \item 
        {\bf $\algo{Verify}\big(\rho, x,\pi\big)$:} On input a string $\rho$, a
        statement $x$, and a proof $\pi$, output either 1 (accepting) or 0 (rejecting).
    \end{itemize}
The following requirements are satisfied:
    \begin{enumerate}
        \item {\bf Completeness:} For every $x \in L$,
        every $\widetilde{\ell} \in \Naturals$, 
        every $\widetilde{D} \ge \widetilde{d}(\abs{x},\widetilde{\ell})$, 
        and every
        $\lambda \in \Naturals$, it holds that
        $$\Pr[ \rho \gets \algo{V}(1^{\lambda}, 1^{\abs{x}},1^{\widetilde{\ell}}, 1^{\widetilde{D}});\pi \gets \algo{P}(\rho,x,w) : \algo{Verify}\big(\rho,x ,\pi\big)=1 ] = 1.$$  
        
        \item {\bf Public coin:}
            $\algo{V}(1^{\lambda}, 1^n,1^{\widetilde{\ell}}, 1^{\widetilde{D}})$ simply outputs a uniformly
            random string.
        
        \item \label{item:def:zap:soundness}
        {\bf Selective non-witness adaptive-statement soundness:} 
        For any non-uniform QPT machine $P^*_{\lambda}$, any $n,\widetilde{D} \in \Naturals$, and any non-witness $\widetilde{w} \in \bits^{*}$,
        \begin{equation*}
            \Pr[
            \begin{array}{l}
            \rho \gets \algo{V}(1^{\lambda}, 1^{n},1^{\abs{\widetilde{w}}}, 1^{\widetilde{D}});\\
            \big(x,\pi^*\big) \gets P^*_{\lambda}(\rho)
            \end{array}: 
            \begin{array}{l}
                \algo{Verify}(\rho,x,\pi^*) = 1 ~\wedge\\
                \widetilde{D} \ge \widetilde{d}(\abs{x}, \abs{\widetilde{w}}) ~\wedge~ (x,\widetilde{w}) \in \widetilde{R}
            \end{array} ] \le \negl(\SecPar).
        \end{equation*}
        
        
        \item {\bf Statistical witness indistinguishability:} For every (possibly unbounded) ``cheating''
        verifier $V^{*}=(V^{*}_{1}, V^{*}_{2})$ and every $n,\widetilde{\ell},\widetilde{D} \in \Naturals$, the
        probabilities
        $$\Pr[V^{*}_{2}(\rho,x,\pi,\zeta)=1 ~\wedge~ (x,w) \in  \mathcal{R} ~\wedge~ (x,w')  \in \mathcal{R} ]$$ 
        in the
        following two experiments differ only by $\negl(\lambda)$:
        \begin{itemize}
            \item {\em Experiment 1:}
            $(\rho,x,w,w',\zeta) \gets
            V^{*}_{1}(1^{\lambda}, 1^{n},1^{\widetilde{\ell}}, 1^{\widetilde{D}})$, $\pi \gets \algo{P}(\rho,x,w)$;
            
                \item {\em Experiment 2:}
            $(\rho,x,w,w',\zeta) \gets
            V^{*}_{1}(1^{\lambda}, 1^{n},1^{\widetilde{\ell}}, 1^{\widetilde{D}}),$ $\pi \gets \algo{P}(\rho,x,w')$.               
        \end{itemize}  
    \end{enumerate}
\end{definition}
\begin{lemma}[\cite{C:CGHKLMPS21}]\label{lem:ZAP:from:QLWE}
Assuming QLWE, there exist ZAPs as per \Cref{def:generalizedzap-interface} for any super-complement language as per \Cref{def:super-complement-language}.
\end{lemma}

\subsection{Construction}
\label{sec:ring-sig:construction}
Our construction $\RS$, shown in \Cref{constr:pq:ring-sig}, relies on the following building blocks: 
\begin{enumerate}
\item
Pair-wise independent hash functions; 
\item
A blind-unforgeable signature scheme $\algo{Sig}$ satisfying \Cref{def:classical-BU-sig}; 
\item
A lossy PKE scheme $\LE$ satisfying \Cref{def:special-LE}; 
\item
A ZAP for special super-complement languages $\algo{ZAP}$ satisfying \Cref{def:generalizedzap-interface}. 
\end{enumerate}

We remark that the $\RS.\Sign$ algorithm runs $\ZAP$ on a special super-complement language $(L, \tilde{L})$, whose definition will appear after the construction in \Cref{sec:building-blocks:language}. This arrangement is because we believe that the language $(L, \tilde{L})$ will become easier to understand once the reader has slight familiarity with \Cref{constr:pq:ring-sig}.


\begin{ConstructionBox}[label={constr:pq:ring-sig}]{Post-Quantum Ring Signatures}
Let $\widetilde{D} = \widetilde{D}(\lambda, N)$ 
be the maximum depth of the $\algo{NP}$ verifier circuit for language $\widetilde{L}$ restricted to statements where the the ring has at most $N$ members, and the security parameter for \Sig and $\algo{LE}$ is $\lambda$. Let $n = n(\lambda,\log N)$ denote the maximum size of the statements of language $L$ where the ring has at most $N$ members and the security parameter is $\lambda$. Recall that for security parameter $\lambda$, secret keys in $\LE$ have size $\widetilde{\ell} = \ell_{\sk}(\lambda)$. We now describe our ring signature construction: 

\subpara{Key Generation Algorithm $\algo{Gen}(1^\lambda,N)$:} 
    \begin{itemize}
        \item sample signing/verification key pair: $(vk,sk) \gets \algo{Sig.Gen}(1^\lambda)$;
        \item sample obliviously an injective public key of $\algo{LE}$: $pk \gets \LE.\KSam^\ls(1^\secpar)$;     
        \item compute the first message $\rho \leftarrow \algo{ZAP.V}(1^\lambda,1^n,1^{\widetilde{\ell}},1^{\widetilde{D}})$ for \ZAP;
        \item output the verification key $\VK\coloneqq (vk,pk,\rho)$ and signing key $\SK\coloneqq (sk,vk,pk,\rho)$.
    \end{itemize}

\subpara{Signing Algorithm $\algo{Sign}(\SK,\Ring, m)$:}
    \begin{itemize}
    	\item parse $\Ring = (\VK_1,\dots,\VK_{\ell})$; and parse $\SK=(sk,vk,pk,\rho)$;
        \item compute $\sigma \gets \algo{Sig.Sign}(sk,\Ring\|m)$; 
        \item let $\VK \coloneqq \VK_i \in R$ be the verification key corresponding to $\SK$; 

        
        \item 
        sample two pairwise-independent hash functions $\PI_1$ and $\PI_2$, and compute
         $$r_{c_1} = \PI_1(\Ring\|m), ~~r_{c_2} = \PI_2(\Ring\|m).$$
        \item compute $c_1 \gets \algo{LE.Enc}(pk,(\sigma,vk);r_{c_1})$ and  $c_2 \gets \algo{LE.Enc}(pk,0^{|\sigma|+|vk|};r_{c_2})$;
        
        \item 
        let $\VK_1 = (vk_1,pk_1,\rho_1)$ denote the lexicographically smallest member of $\Ring$ (as a string; note that this is necessarily unique); 
        
        \item 
        fix statement $x=(\Ring, m, c_1,c_2)$ and witness $w=(vk,pk, \sigma,r_{c_1})$. We remark that this statement and witness correspond to a super-complement language $(L, \tilde{L})$ that will be defined in \Cref{sec:building-blocks:language}. Looking ahead, $x$ with witness $w$ is a statement in the $L$ defined in \Cref{eq:def:language:L}; $x$ constitutes a statement that is {\em not} in the $\tilde{L}$ defined in \Cref{eq:def:language:L-tilde}. 
        \item sample another pairwise-independent hash function $\PI_3$ and compute $r_\pi = \PI_3(\Ring\|m)$;
        \item compute $\pi \gets \algo{ZAP.P}(\rho_1,x,w; r_\pi)$;
        \item output $\Sigma = (c_1,c_2,\pi)$.
    \end{itemize}

\subpara{Verification Algorithm $\algo{Verify}(\Ring,m, \Sigma)$:}
    \begin{itemize}
        \item identify the lexicographically smallest verification key $\VK_1$ in $\Ring$;
       
        \item fix $x=(\Ring, m, c_1, c_2)$;  read $\rho_1$ from $\VK_1$; 
        \item compute and output $\algo{ZAP}.\algo{Verify}(\rho_1,x,\pi)$.
\end{itemize}
\end{ConstructionBox}

\subsubsection{The Super-Complement Language Proven by the ZAP}
\label{sec:building-blocks:language}
We now define the super-complement language $(L, \widetilde{L})$ used in \Cref{constr:pq:ring-sig}. This deviates from the $(L, \tilde{L})$ defined in \cite[Section 5]{C:CGHKLMPS21}, to accommodate \Cref{constr:pq:ring-sig}.  

For a statement of the form $x_1 = (\Ring, m,c)$ and witness $w =  \big(\mathsf{VK}=(vk,pk,\rho),\sigma,r_c \big)$, define relations $R_1$, $R_2$, and $R_3$ as follows:
\begin{align*}
	(x_1,w) \in R_1 &  ~\Leftrightarrow~ \VK \in \Ring, \\
    (x_1,w) \in R_2 &  ~\Leftrightarrow~ \algo{LE.Enc}\big(pk,(\sigma,vk);r_c\big) = c,\\
    (x_1,w) \in R_3 & ~\Leftrightarrow~ \algo{Sig.Verify}(vk, \Ring\|m,\sigma)=1. 
\end{align*}
Next, define the relation $R'$ as  $R'\coloneqq R_1\cap R_2\cap R_3$. Let $L'$ be the language corresponding to $R'$. Define language $L$ as 
\begin{equation}\label{eq:def:language:L}
L \coloneqq \big\{x=(\Ring, m,c_1,c_2) ~\big|~ 
(\Ring, m, c_1) \in L' ~\vee~ 
(\Ring, m, c_2) \in L' 
\big\}.
\end{equation}

Now, we define another language $\widetilde{L}$ and prove that it is a super-complement of $L$ in \Cref{lemma:ring-sig:complement}. 
Let $x_1 = (\Ring, m,c)$ as above, but let $\widetilde{w} \coloneqq msk$. Define the following relations:
\begin{align}
    (x_1,\widetilde{w})\in R_4 & ~\Leftrightarrow~ \forall j \in [\ell]:  \algo{LE.Valid}\big(pk_j, \algo{LE.MSKExt}(msk, pk_j)\big)=1 
    \label{def:relation:R4}\\
    (x_1,\widetilde{w})\in R_5 & ~\Leftrightarrow~ 
    \left\{
        \begin{array}{l}
        \exists\algo{VK} \in \mathsf{R}: \algo{VK} = (vk,pk,\rho) ~\text{such that:}\\
         \algo{LE.Valid}\big(pk, \algo{LE.MSKExt}(msk, pk)\big)=1 ~\wedge\\
        \algo{LE.Dec}\big(\algo{LE.MSKExt}(msk, pk), c \big) = (\sigma,vk) ~\wedge\\ 
        \algo{Sig.Verify}(vk, \Ring\|m, \sigma)=1 
        \end{array} 
    \right. \label{def:relation:R5}
\end{align}
where, for each $j \in [\ell]$,  $\algo{VK}_j = (vk_j, pk_j, \rho_j)$ is the $j$-th member in $\Ring$.  Let $L_4$ and $L_5$ be the languages corresponding to $R_4$ and $R_5$, respectively. Define further the relation $\widehat{R}$ according to $\widehat{R} \coloneqq R_4 \setminus R_5$, and let $\widehat{L}$ be the corresponding language. Define $\tilde{L}$ as follows:
\begin{equation}\label{eq:def:language:L-tilde}
\widetilde{L} \coloneqq \big\{x=(\Ring, m,c_1,c_2) ~\big|~ 
(\Ring, m,c_1) \in \widehat{L}  ~\wedge~
(\Ring, m,c_2) \in \widehat{L}
\big\}.
\end{equation}

Following a similar proof as for \cite[Lemma 5.1]{C:CGHKLMPS21}, we can show that $\tilde{L}$ is indeed a super-complement of $L$.
\begin{MyClaim}
    \label{lemma:ring-sig:complement}
    If $\LE$ satisfies the completeness defined in \Cref{item:def:LE:completeness} of \Cref{def:special-LE}, then the language $\widetilde{L}$ defined in \Cref{eq:def:language:L-tilde} is a super-complement (as per \Cref{def:super-complement-language}) of the language $L$ defined in \Cref{eq:def:language:L}.
\end{MyClaim}
\begin{proof}
To prove this claim, we need to show that for any statement $x$ of the following form 
\begin{equation}\label{eq:ring-sig:complement:statement:x}
x =  (\Ring, m, c_1, c_2),
\end{equation}
it holds that $x \in \tilde{L} \Rightarrow x \notin L$ (see \Cref{rmk:super-complement-language:special-case}). In the following, we finish the proof by showing the contrapositive: $x \in L \Rightarrow x\notin \tilde{L}$. 

For any $x$ as in \Cref{eq:ring-sig:complement:statement:x}, we define 
$$x_1 \coloneqq (\Ring, m,c_1) ~\text{and}~ x_2 \coloneqq (\Ring, m,c_2).$$
To prove ``$x \in L \Rightarrow x\notin \tilde{L}$'', it suffices to show that the following \Cref{eq:ring-sig:complement:expression1,eq:ring-sig:complement:expression2} hold for every $w= (\algo{VK}= (vk,pk,\rho),\sigma,r_c)$ and every $\widetilde{w} = msk$:
\begin{align}
&(x_1,w) \in R' \wedge \big(x_1,\widetilde{w}\big) \in R_4 ~\Rightarrow~ \big(x_1, \widetilde{w} \big) \in R_5 \label[Expression]{eq:ring-sig:complement:expression1}\\
&(x_2,w) \in R' \land \big(x_2,\widetilde{w}\big) \in R_4 ~\Rightarrow~ \big(x_2, \widetilde{w} \big) \in R_5. \label[Expression]{eq:ring-sig:complement:expression2}
\end{align}

We first prove \Cref{eq:ring-sig:complement:expression1}. If $(x_1,\tilde{w}) \in R_4$, then for all $\algo{VK} = (vk, pk, \rho) \in \algo{R}$, we know that $\algo{LE.MSKExt}(msk, pk)$ is a valid secret key for $pk$.
This means that:
\begin{itemize}
\item[] {\bf Fact:} any ciphertext w.r.t.\ any $pk$ (contained in any $\VK$) in $\Ring$ can be decrypted correctly by $\algo{LE.MSKExt}(msk, pk)$. 
\end{itemize} 
Also, observe that $(x_1,\tilde{w}) \in R'$ means $(x_1,w) \in R_1 \cap R_2 \cap R_3$, which says $c_1$ is a valid ciphertext of a signature for $\Ring\|m$, encrypted by some $pk$ in the ring $\Ring$. Then,  by the above {\bf Fact}, we must have $(x_1, \tilde{w}) \in R_5$.


\Cref{eq:ring-sig:complement:expression2} can be proven similarly. This finish the proof of \Cref{lemma:ring-sig:complement}.
\end{proof}
\subsection{Proof of Security} 
We now prove that \Cref{constr:pq:ring-sig} is a post-quantum secure post-quantum ring signature satisfying \Cref{def:pq:ring-signatures}. Its completeness follows straightforwardly from the completeness of $\ZAP$ and $\algo{Sig}$. We next prove post-quantum anonymity and blind-unforgeability in \Cref{sec:prove:pq:anonymity} and \Cref{sec:prove:pq:blind-unforgeability}, respectively.



\subsubsection{Proving Post-Quantum Anonymity}
\label{sec:prove:pq:anonymity}

In this section, we prove the following \Cref{lem:ring-sig:pq:anonymity}, which establishes post-quantum anonymity for \Cref{constr:pq:ring-sig}.
\begin{lemma}
\label{lem:ring-sig:pq:anonymity}
    Assume $\algo{LE}$ satisfies the lossiness (\Cref{def:LE:property:lossyness}) described in \Cref{def:special-LE} and $\algo{ZAP}$ is statistically witness indistinguishable. Then, \Cref{constr:pq:ring-sig} satisfies the post-quantum anonymity described in \Cref{def:pq:anonymity}.
\end{lemma}

Let $\Adv$ be a QPT adversary participating in \Cref{expr:pq:anonymity}. Recall that the classical identities specified by $\Adv$ is $(i_0, i_1)$ and the quantum query is $\sum_{\Ring, m, t} \psi_{\Ring, m, t} \ket{\Ring, m, t}$. We will show a sequence of hybrids where the challenger switches from signing using $i_0$ to signing using $i_1$. It is easy to see that the scheme is post-quantum anonymous if $\Adv$ cannot tell the difference between each pair of adjacent hybrids.

\para{Hybrid $H_0$:} This hybrid simply runs the anonymity game with $b=0$. That is, $\Adv$'s query is answered as follows:
$$\sum_{\Ring, m, t}\psi_{\Ring, m, t} \ket{ \Ring, m, t} \mapsto \sum_{\Ring, m, t}\psi_{\Ring, m, t} \ket{\Ring, m, t\xor f(\Ring, m)},$$
where
$f(\Ring, m) \coloneqq
\begin{cases}
\RS.\Sign(\SK_{i_0}, \Ring\| m;r) & \text{if}~\VK_{i_0}, \VK_{i_1} \in R \\
\bot & \text{otherwise}
\end{cases}
$. We remark that $f(\Ring,m)$ is performed quantumly for each $(\Ring, m)$ pair in the superposition. We say that $\Adv$ {\em wins} if it outputs $b' = b$ $(= 0)$.

It is worth noting that although $\RS.\Sign$ is a randomized algorithm, it uses only a single random tape $r$ for all the $(\Ring, m)$ pairs in the superposition (See \Cref{rmk:single-randomness}). In \Cref{constr:pq:ring-sig}, this means that the pair-wise independent hash functions $\PI_1, \PI_2, \PI_3)$ are sampled only once (i.e., they remain the same for all the $(\Ring, m)$ pairs in the superposition).

\para{Hybrid $H_1$:} In this hybrid, for each signing query from $\Adv$, instead of sampling a pair-wise indepedent function $\PI_2(\cdot)$ and compute $r_{c_2} = \PI_2(\Ring\|m)$, we  compute $r_{c_2} = P_2(\Ring\|m)$, where $P_2(\cdot)$ is a {\em random} function. In effect, $r_{c_2}$ is now randomly sampled for each $(\Ring, m)$ pairs.  

\subpara{$H_0 \idind H_1$:} 
This follows from \Cref{lemma:2qwise}. 

\para{Hybrid $H_2$:} Here we switch $c_2$ from an encryption of a zero string to $c_2 \gets \algo{LE.Enc}(pk_{i_1},(\sigma',vk_{i_1});r_{c_2})$, where $\sigma' \gets \Sig.\Sign(sk_{i_1}, \Ring\|m)$. In this hybrid, it is worth noting that the previous ``dummy ciphertext'' $c_2$ becomes a valid one, i.e., it encrypts a valid signature for $\Ring\|m$ using identity $i_1$.


\subpara{$H_1 \sind H_2$:} In both $H_1$ and $H_2$, $r_{c_2}$ is sampled (effectively) uniformly at random for each $(\Ring, m)$ pair in the superposition in each signing query. Consider an oracle $\mathcal{O}$ that takes $(\Ring, m)$ as input and returns $(c_1,c_2,\pi)$ just as in $H_1$, and an analogous oracle $\mathcal{O}'$ that takes the same input and returns $(c_1,c_2,\pi)$ computed just as in $H_2$. Note that the only difference between the outputs of $\mathcal{O}$ and $\mathcal{O}'$ is in $c_2$, which encrypts $0^{|\sigma|+|vk|}$ in $H_1$ and $(\sigma',vk_{i_1})$ in $H_2$. Recall that $pk_{i_1}$ is produced using $\LE.\KSam^\ls$ and therefore, by lossiness (\Cref{def:LE:property:lossyness}), we have that the distributions of $c_2$ in $H_1$ and $H_2$ are statistically indistinguishable, implying that the outputs of $\mathcal{O}$ and $\mathcal{O}'$ are statistically close for every input $(\Ring,m)$, say less than distance $\Delta$ (which is negligible in \secpar). Then, by \Cref{lemma:oracleindist}, the probability that $\Adv$ distinguishes these two oracles even with $q = \poly(\secpar)$ quantum queries is at most $\sqrt{8C_0q^3\Delta}$, which is negligible since $\Delta$ is negligible. Similarity of these hybrids is immediate.

\para{Hybrid $H_3$:} In this hybrid, we switch back to using the pairwise independent hash function $\PI_2$ to compute $r_{c_2}$, instead of using a truly random function. Effectively we are undoing the change made in $H_1$. 

\subpara{$H_2 \idind H_3$:} This again follows from \Cref{lemma:2qwise}.

\para{Hybrid $H_4$:} Here, we compute $r_\pi$ as the output of a {\em random} function $r_\pi = P_3(\Ring\|m)$, instead of being computed using $\PI_3$ as before. In effect, $r_\pi$ is now uniformly random. 

\subpara{$H_3 \idind H_4$:} This again follows from \Cref{lemma:2qwise}.

\para{Hybrid $H_5$:} As mentioned in $H_2$, the ``block'' $(\Ring, m, c_2)$ is valid. Recall that in previous hybrids, $\ZAP$ uses the witness $w$ corresponding to the block $(\Ring, m, c_1)$. In this hybrid, we switch the witness used by $\ZAP$ from $w=(vk_{i_0},pk_{i_0}, \sigma, r_{c_1})$ to $w'=(vk_{i_1},pk_{i_1}, \sigma', r_{c_2})$, i.e., the witness corresponding to the $(\Ring, m, c_2)$ block. 

\subpara{$H_4 \sind H_5$:} In both $H_4$ and $H_5$, $r_\pi$ is sampled (effectively) uniformly at random for each $(\Ring, m)$ pairs in the superposition for each query. 
Consider an oracle $\mathcal{O}$ that takes $(\Ring,m)$ as input and returns $(c_1,c_2,\pi)$ just as in $H_4$, and an analogous oracle $\mathcal{O}'$ that takes the same input and returns $(c_1,c_2,\pi)$ computed just as in $H_5$. Note that the only difference between the outputs of $\mathcal{O}$ and $\mathcal{O}'$ is in $\pi$, which is generated using $w$ in $H_4$ and using $w'$ in $H_5$. Since both $w$ and $w'$ are valid witnesses, by the {\em statistical} witness indistinguishability of $\algo{ZAP}$, we have that the distributions of $\pi$ in $H_4$ and $H_5$ are statistically indistinguishable for every $(\Ring, m)$ pair (aka the input to the $\mathcal{O}$ or $\mathcal{O}'$). In other words, the outputs of $\mathcal{O}$ and $\mathcal{O}'$ are statistically close for every input $(\Ring,m)$, say less than distance $\Delta$ (which is negligible in \secpar). Then, the statistical indistinguishability follows from \Cref{lemma:oracleindist}.

\para{Hybrid $H_6$:} In this hybrid, we switch back to using the pairwise independent hash function $\PI_3$ to compute $r_\pi$, instead of using a truly random function. Effectively we are undoing the change made in $H_4$.

\subpara{$H_5 \idind H_6$:} This again follows from \Cref{lemma:2qwise}.

\para{Hybrid $H_7$:} In this hybrid, instead of sampling $r_{c_1} = \PI_1(\Ring\|m)$, we instead compute $r_{c_1}$ as the output of a {\em random} function $r_{c_1} = P_1(\Ring\|m)$. In effect, $r_{c_1}$ is now randomly sampled.

\subpara{$H_6 \idind H_7$:} This again follows from \Cref{lemma:2qwise}.

\para{Hybrid $H_8$:} In this hybrid, we switch $c_1$ from an encryption of $(\sigma, vk_{i_0})$ to one of $(\sigma', vk_{i_1})$.

\subpara{$H_7 \sind H_8$:} This follows from the same argument for $H_1\sind H_2$.

\para{Hybrid $H_9$:} In this hybrid, we switch back to using the pairwise independent hash function $\PI_1$ to compute $r_{c_1}$, instead of using a truly random function. Effectively we are undoing the change made in $H_7$.

\subpara{$H_8 \idind H_9$:} This again follows from \Cref{lemma:2qwise}.



\para{Hybrid $H_{10}$:} In this hybrid, we switch to computing $r_\pi$ as the output of a {\em random} function $r_\pi = P_3(\Ring\|m)$, instead of being computed using $\PI_3$.  

\subpara{$H_9 \idind H_{10}$:} This again follows from \Cref{lemma:2qwise}.

\para{Hybrid $H_{11}$:} In this hybrid, we again switch the witness used to generate $\pi$, from $w'$ to $w''=(vk_{i_1},pk_{i_1}, \sigma',r_{c_1})$.
 
\subpara{$H_{10} \sind H_{11}$:} This follows from the same argument for $H_4 \sind H_5$.

\para{Hybrid $H_{12}$:} In this hybrid, we switch back to using the pairwise independent hash function $\PI_3$ to compute $r_\pi$, instead of using a truly random function. 

\subpara{$H_{11} \idind H_{12}$:} This again follows from \Cref{lemma:2qwise}.

\para{Hybrid $H_{13}$:} Here, we switch to computing $r_{c_2}$ as the output of a {\em random} function $r_{c_2} = P_2(\Ring\|m)$.

\subpara{$H_{12} \idind H_{13}$:}  This again follows from \Cref{lemma:2qwise}.  

\para{Hybrid $H_{14}$:} In this hybrid, we switch $c_2$ to an encryption of zeroes, namely $c_2 = \LE.\Enc(pk, 0^{|\sigma|+|vk|}; r_{c_2})$, instead of an encryption of $(\sigma',vk_{i_1})$.

\subpara{$H_{13} \sind H_{14}$:} This argument is identical to that for simlarity between $H_1 \sind H_2$.

\para{Hybrid $H_{15}$:} In this hybrid, we switch back to using the pairwise independent hash function $\PI_2$ to compute $r_{c_2}$, instead of using a truly random function. 

\subpara{$H_{14} \idind H_{15}$:} This again follows from \Cref{lemma:2qwise}.  


\vspace{1em}

Observe that $H_{15}$ corresponds to sign using identity $i_1$ in \Cref{expr:pq:anonymity}. This finishes the proof of \Cref{lem:ring-sig:pq:anonymity}.

\subsubsection{Proving Post-Quantum Blind-Unforgeability}
\label{sec:prove:pq:blind-unforgeability}

In this section, we prove the following \Cref{lem:ring-sig:pq:blind-unforgeability}, which establishes post-quantum blind-unforgeability for \Cref{constr:pq:ring-sig}.
\begin{lemma}
\label{lem:ring-sig:pq:blind-unforgeability}
    Assume $\Sig$ is blind-unforgeable as per \Cref{def:classical-BU-sig}, $\algo{LE}$ satisfies the completeness of master secret keys property (\Cref{def:LE:property:GenWithSK:completeness}) and the almost-unique secret key property (\Cref{item:def:LE:computationally-unique-sk}), and $\algo{ZAP}$ has the selective non-witness adaptive-statement soundness (\Cref{item:def:zap:soundness}). Then, \Cref{constr:pq:ring-sig} is blind-unforgeable as  per \Cref{def:pq:blind-unforgeability}.
\end{lemma}

Consider a QPT adversary $\Adv_\RS$ participating in \Cref{expr:pq:blind-unforgeability}. We proceed using a sequence of hybrids to set up our reduction to the blind-unforgeability of \algo{Sig}.
    
\para{Hybrid $H_0$:} This is just the post-quantum blind-sunforgeability game (\Cref{expr:pq:blind-unforgeability}) for our construction. In particular, for all $i\in [Q]$, the encryption key $pk_i$ is generated as $pk_i\gets \LE.\KSam^\ls(1^\secpar;r_i)$. Recall that we are in the full key exposure setting, so both the public keys and random coins $\Set{pk_i, r_i}_{i \in [Q]}$ are given to $\Adv$.   

\para{Hybrid $H_1$:} In this experiment, the only difference is that, the challenger generates the $\Set{pk_i}_{i \in [Q]}$ by running $\big( \Set{pk_i}_{i\in [Q]}, \msk \big)\gets \algo{LE.MSKGen}(1^\secpar, Q)$. The challenger keeps $\msk$ to itself, and sends $\big\{pk_i, \algo{LE.RndExt}(pk_i)\big\}_{i\in [Q]}$ to $\Adv$.

\subpara{$H_0 \cind H_1$:} This follows immediately from the IND of $\algo{MSKGen}$/$\KSam^\ls$ property (\Cref{def:LE:property:IND:GenWithSK-KSam}) of $\LE$ as specified in \Cref{def:special-LE}. It is worth noting that $\Adv_\RS$'s quantum access to the signing algorithm does not affect this proof at all, since the $pk_i$'s (contained in $\VK_i$'s) are sampled classically by the challenger before $\Adv_\RS$ makes any quantum {\sf sign} queries.

\para{Reduction to the BU of $\Sig$.} We proceed to show that post-quantum blind-unforgeability holds in $H_1$. Consider the adversary's forgery attempt 
$$\big(\algo{R}^*, m^*, \Sigma^* = (c_1^*, c_2^*, \pi^*)\big)~\text{satisfying}~(\Ring^*, m^*) \in B^\RS_\epsilon.$$ 
Let $x^* \coloneqq (\Ring^*, m^*, c^*_1, c^*_2)$.  Let $\algo{VK}_1^*$ = ($vk_1^*,pk_1^*,\rho_1^*$) be the lexicographically smallest verification key in $\algo{R}^*$.



Observe that for the $x^*$ defined above, one of the following two cases must happen: $x^* \in \tilde{L}$ or $x^* \notin \tilde{L}$. (Recall that $\tilde{L}$ is the super-complement of $L$ defined in \Cref{eq:def:language:L-tilde}.) In the following, we show two claims. \Cref{lemma:unfroge:inltilde} says that it cannot be the case that $x^* \in \tilde{L}$, unless the $\ZAP$ verification rejects (thus, the forgery is invalid). \Cref{lemma:unforge:sig} says that $x^* \notin \tilde{L}$ cannot happen either. Therefore, \Cref{lemma:unfroge:inltilde,lemma:unforge:sig} together show that any QPT adversary has negligible chance of winning the blind-unforgeability game for $\RS$ in $H_1$. Note that winning the post-quantum blind-unforgeability game for $\RS$ is an event that can be efficiently tested. Thus, by $H_0 \cind H_1$, no QPT adversaries can win the post-quantum blind-unforgeability game for $\RS$ in hybrid $H_0$. This concludes the proof of \Cref{lem:ring-sig:pq:blind-unforgeability}.

Now, the only thing left is to state and prove \Cref{lemma:unfroge:inltilde,lemma:unforge:sig}, which is done in the following.

 \begin{MyClaim}
    	\label{lemma:unfroge:inltilde}
    	In $H_1$, assume that $\algo{ZAP}$ satisfies selective non-witness adaptive statement soundness (\Cref{item:def:zap:soundness}). 
    	Then, the following holds:
    	$$\Pr[x^* \in \widetilde{L}  ~\wedge~ \algo{\algo{ZAP}.Verify}(\rho_1^*,x^*,\pi^*)=1] = \negl(\lambda).$$
    \end{MyClaim}   
\begin{proof}
First, notice that by definition, the $\Ring^*$ in $\Adv_\RS$'s forgery contains only $\VK$'s from the set $\mathcal{VK} = \Set{\VK_j}_{j\in [Q]}$ generated by the challenger. Therefore, it suffices to show that for each $j \in [Q]$, 
\begin{equation}\label{eq:lemma:unfroge:inltilde:real-goal}
\Pr[x^*  \in \widetilde{L} ~\wedge~ \algo{ZAP.Verify}(\rho_j,x^*,\pi^*)=1] = \negl(\lambda),
\end{equation}
where $\rho_j$ denotes the first-round message of \ZAP corresponding to the $j$-th verification key $\algo{VK}_j$ generated in the game.

Let $\Adv_\RS$ be an adversary attempting to output a forgery such that 
$$x^* \in \widetilde{L}~~\text{and}~~\algo{ZAP}.\Verify(\rho_j,x^*,\pi^*) =1.$$
 We build an adversary $\Adv_\ZAP$ against the selective non-witness adaptive-statement soundness of $\algo{ZAP}$ for $(L, \widetilde{L})$ (defined in \Cref{eq:def:language:L,eq:def:language:L-tilde} respectively). 
The algorithm $\Adv_\ZAP$ proceeds as follows: 
    	\begin{itemize}
    		\item On input the 1st \algo{ZAP} message $\widehat{\rho}$, it sets $\rho_{j} = \widehat{\rho}$ and  proceeds exactly as $H_1$. 
    		\item Upon receiving the forgery attempt $\big(\Ring^*, m^*, \Sigma^* =(c^*_1, c^*_2, \pi^*)\big)$ from $\Adv$, it   outputs $$\big(x^* \coloneqq (\Ring^*, m^*, c^*_1, c^*_2),~\pi^*\big).$$
    	\end{itemize} 

We remark that $H_1$ is quantum. So, $\Adv_\ZAP$ needs to be a quantum machine to simulate $H_1$ for $\Adv_\RS$. This is fine since we assume that the soundness (\Cref{item:def:zap:soundness}) of ZAP in \Cref{def:generalizedzap-interface} holds against QPT adversaries. 

To finish the proof, notice that $x^* \in \widetilde{L}$ means that there exists a ``non-witness'' $\tilde{w}^*$ such that $(x^*, \tilde{w}^*) \in \tilde{R}$. Therefore, if \Cref{eq:lemma:unfroge:inltilde:real-goal} does not hold, $(\rho_j, x^*, \pi^*)$ will break the soundness (\Cref{item:def:zap:soundness}) w.r.t.\ the non-witness $\tilde{w}$.
\end{proof}

\begin{MyClaim}
	\label{lemma:unforge:sig}
	In $H_1$, assume that $\algo{Sig}$ satisfies the blind-unforgeability as per \Cref{def:classical-BU-sig}, $\algo{LE}$ satisfies the completeness of master secret keys propoerty (\Cref{def:LE:property:GenWithSK:completeness}) and the almost-unique secret key property (\Cref{item:def:LE:computationally-unique-sk}). Then,
	$\Pr\big[x^* \not \in \widetilde{L}\big] = \negl(\lambda)$.
\end{MyClaim}
\begin{proof}
Let $\Adv_\RS$ be a QPT adversary attempting to output a forgery w.r.t.\ our $\RS$ scheme such that $x^* \not \in \widetilde{L}$. We build an algorithm $\Adv_\Sig$ against the  blind-unforgeability of $\algo{Sig}$. The algorithm $\Adv_\Sig$ (playing the blind-unforgeability game \Cref{expr:BU:ordinary-signature} for $\Sig$) proceeds as follows:
\begin{enumerate}
	\item 
	invoke $\Adv_\RS$ to obtain the $\epsilon$ for the blind-unforgeability game of $\RS$; give this $\epsilon$ to $\Adv_\Sig$'s own challenger for the blind-unforgeability game of $\Sig$;
	
	\item
	receive $\widehat{vk}$ from its own challenger; pick an index $j \pick [Q]$ uniformly at random; set $vk_{j} \coloneqq \widehat{vk}$; then, proceeds as in $H_1$ to prepare the rest of the verification keys and continue the execution with $\Adv_\RS$.
	
	\item \label[Step]{step:unforge:sig:signing-query}
	when $\Adv_\RS$ sends a quantum signing query $(\algo{sign},i, \sum \psi_{\Ring, m, t}\ket{\Ring, m, t})$, if the specified identity $i$ is not equal to $j$, it proceeds as in $H_1$; otherwise, it uses the blind-unforgeability (for $\Sig$) game's signing oracle $\Sig.\Sign$ to obtain a $\Sig$ signature for the $j$-th party and then continues exactly as in $H_1$; (See the paragraph right after the description of $\Adv_\Sig$.)

	\item 
	if $\Adv_\RS$ tries to corrupt the $j$-th party, $\Adv_\Sig$ aborts; (It is worth noting that the identities are classical. So, $\Adv_\RS$'s quantum power does not affect this step.)

	\item 
	upon receiving the forgery attempt $\Sigma^*$ from $\Adv_\RS$, $\Adv_\Sig$ decrypts $c_1^*$ using $\msk$ to recover $\sigma_1^*$. (Recall that, the secret key for $pk_j$ can be obtained as $\algo{LE.MSKExt}(\msk, pk_j)$).  If  $$\algo{Sig.Verify}(vk_j, \Ring^*\|m^*, \sigma_1^*)=1,$$ it sets $\widehat{\sigma} := \sigma_1^*$. Otherwise, it decrypts $c_2^*$ with $\msk$ to recover $\sigma_2^*$, and sets $\widehat{\sigma} \coloneqq \sigma_2^*$. It outputs $(\Ring^*\|m^*,\widehat{\sigma})$. 
\end{enumerate}

We first remark that, up to (inclusively) \Cref{step:unforge:sig:signing-query}, $\Adv_\RS$'s view is identical to that in $H_1$. Recall that in $H_1$, the challenger maintains a blindset $B^\RS_\epsilon$ such that any $(\Ring, m) \in B^\RS_\epsilon$ will not be answered (this is inherited from $H_0$, which is exactly \Cref{expr:pq:blind-unforgeability}). In contrast, in the execution of $\Adv_\Sig$ described above, $\Adv_\Sig$ first forwards the $\sum \psi_{\Ring, m, t}\ket{\Ring, m, t}$ part of $\Adv_\RS$'s query to its $\Sig.\Sign$ oracle to obtain 
$\sum_{\Ring, m, t} \psi_{\Ring, m, t} \ket{\Ring, m, t\xor B^\Sig_\epsilon\Sig.\Sign(sk_j, \Ring\|m)}$ (note that $sk_j = \widehat{sk}$), and then performs the remaining computation exactly as in $H_1$. Note that the $B^\Sig_\epsilon$ is the blindset maintained by the $\Sig$ signing algorithm. Importantly, since the ``messages'' singed by $\Sig$ are of the form $\Ring\|m$, $B^\Sig_\epsilon$ is actually generated identically to $B^\RS_\epsilon$---that is, both of them are generated by including each $(\Ring, m)$ pair in with (the same) probability $\epsilon$.

To finish the proof, we show that 
$(\Ring^*\|m^*,\widehat{\sigma})$ is a valid forgery against $\Sig$'s blind-unforgeability game with probability at least $\frac{1}{Q}\big(\Pr\big[x^* \not \in \widetilde{L}\big]- \negl(\lambda)\big)$.

Recall that we are focusing on the case $x^* \notin \tilde{L}$, where $\tilde{L}$ is defined in \Cref{eq:def:language:L-tilde}; without loss of generality, assume that $(\Ring^*, m^*, c_1^*) \not \in \widehat{L}$. Then, observe that due to the way $H_1$ generates the public keys (more acurately, \Cref{def:LE:property:GenWithSK:completeness}) and that $\Ring^* \subseteq \mathcal{VK}\setminus \mathcal{C}$ (in particular, $\Ring^*\subseteq \mathcal{VK}$), we have 
    	\begin{equation}\label[Expression]{eq:unforge:sig:proof:L4}
    	\big((\Ring^*, m^*,c_1^*), \msk\big) \in R_4 ~~\text{(recall that $R_4$ is defined in \Cref{def:relation:R4})}.
    	\end{equation}
Since we assume that $(\Ring^*, m^*,c_1^*) \notin \widehat{L}$, \Cref{eq:unforge:sig:proof:L4} and the definition of $\widehat{L}$ imply the existence of a string $\widetilde{w}$ such that
\begin{equation}\label[Expression]{eq:unforge:sig:proof:L5}
\big((\Ring^*, m^*,c_1^*),\widetilde{w}\big)  \in R_5 ~~\text{(recall that $R_5$ is defined in \Cref{def:relation:R5})}.
\end{equation}
We remark that the $\tilde{w}$ may not equal $\msk$. However, note that $R_5$ (\Cref{def:relation:R5}) tests if 
$$\algo{LE.Valid}\big(pk, \algo{LE.MSKExt}(\tilde{w},pk)\big) = 1$$ with respect to the $pk$ contained in some $\VK$ in the ring. If this test passes, by \Cref{eq:unforge:sig:proof:L4} and the almost-unique secret key property (\Cref{item:def:LE:computationally-unique-sk}) of $\LE$, it must hold for this $pk$ that $$\algo{LE.MSKExt}(\tilde{w}, pk) = \algo{LE.MSKExt}(\msk, pk),$$ except with negligible probability. 

To summarize, the above argument implies the following facts:
\begin{enumerate}
	\item 
	by our assumption, $(\Ring^*, m^*) \in B^\RS_\secpar$; this also implies $(\Ring^*, m^*) \in B^\Sig_\secpar$ because $B^\Sig_\epsilon = B^\RS_\epsilon$ as argued earlier;


\item 
by \Cref{eq:unforge:sig:proof:L5}, for some $\VK = (vk^*, pk^*, \rho^*) \in \Ring^*$, it must hold  that 
$$\LE.\Dec\big(\algo{LE.MSKExt}(\tilde{w},pk), c_1^*\big) = (\sigma^*,vk^*)~~\text{and}~~\algo{Sig.Verify}(vk^*, \Ring^*\|m^*,\sigma^*)=1.$$ Also, as mentioned earlier, $\algo{LE.MSKExt}(\tilde{w}, pk^*) = \algo{LE.MSKExt}(\msk, pk^*)$ for this $pk^*$.
\end{enumerate}
The above means that the $\Adv_\RS$ uses a
$\algo{VK^*}=(vk^*, pk^*, \rho^*) \in \mathsf{R^*} \subseteq \mathcal{VK}\setminus\mathcal{C}$ such that $c_1^*$ encrypts (among other things) a signature $\sigma^*$ that is valid for the forgery message $\Ring^*\|m^*$ w.r.t. key $vk^*$ (for the blind-unforgeability game of $\Sig$). Moreover, $\Adv_\Sig$ can extract this forgery message efficiently by decrypting $c^*$ using $\algo{LE.MSKExt}(\msk, pk^*)$! 

Finally, observe that index $j$ is sampled uniformly. 
Therefore, we have that $(\widehat{vk} =)$ $vk_j = vk^*$ with probability $1/Q$.
\end{proof}

\subsection{Discussion on Compactness}
\label{sec:discussion}
Our construction of post-quantum ring signatures (i.e., \Cref{constr:pq:ring-sig}) is currently only of theoretical interest. It is not efficient, and it does not enjoy compactness (i.e., the signatures size is independent of, or even poly-logarithmic on, the ring size). It is an interesting problem for future research to construction practical or compact ring signatures that satisfies our notion of post-quantum security. In the following, we briefly discuss why this seems non-trivial.

\para{Efficiency.} Almost all known efficient ring signatures are in the random oracle model, following the Fiat-Shamir paradigm (e.g., \cite{AFRICACRYPT:ABBFG13,EC:LLNW16,torres2018post,DBLP:conf/icics/BaumLO18,DBLP:journals/ijhpcn/WangZZ18,CCS:EZSLL19,AC:BeuKatPin20,DBLP:conf/crypto/LyubashevskyNS21}). Although these constructions are based on post-quantum hardness assumptions, their security proofs can only handle adversaries making {\em classical} random oracle queries. Making these constructions secure in the QROM requires a quantum version of the {\em forking lemma} \cite{EC:PoiSte96,CCS:BelNev06}, which seems hard to prove. Indeed, this problem is still open even for (ordinary) signatures in the post-quantum setting (e.g., see the discussion in \cite{AC:Unruh17}). (As a side note, our construction in \Cref{sec:BU:sig:QROM} does not face this problem as it follows the hash-and-sign paradigm, instead of Fiat-Shamir.)  

\para{Compactness.} The original construction in \cite{C:CGHKLMPS21} does achieve compactness. Although based on their work, our construction in \Cref{sec:PQ-ring-sig:def} gives up compactness by using the underlying $\Sig$ to sign $\Ring \| m$ together\footnote{Note that our construction will not become compact even if we use a compact $\Sig$. This is because the size of the $\ZAP$ proof for the validity of the signature for $\Ring\|m$ also depends on the size of the rings.}; in contrast, \cite{C:CGHKLMPS21} only uses $\Sig$ to sign $m$. Our choice is critical to achieving BU: when proving BU for our $\RS$, we need to reduce to the BU of $\Sig$. The $\RS$ game will ``blind'' $(\Ring, m)$ pairs, while the $\Sig$ game only blinds messages $m$. If we do not use $\Ring\|m$ as the message for $\Sig$ to sign, the reduction will not be able to create the blindset in a consistent manner. This problem cannot be resolved by applying some type of ``hash'' function on $\Ring\|m$ and asking $\Sig$ to sign the short digest. Indeed, blinding $(\Ring, m)$ pairs with probability $\epsilon$ is different from blinding the hash result of $\Ring\|m$, unless the ``hash'' has pseudo-random output. Replacing the ``hash'' with a PRF does not work either, as the verifier also needs to evaluate the ``hash'' to verify the signature. We leave it as an open question to construct {\em compact} ring signatures achieving our post-quantum security notion.

\section{Acknowledgments}
We thank the anonymous PKC 2022 reviewers for their valuable comments.

Rohit Chatterjee and Xiao Liang are supported in part by Omkant Pandey's DARPA SIEVE Award HR00112020026 and NSF grants 1907908 and 2028920. Any opinions, findings, and conclusions, or recommendations expressed in this material are those of the author(s) and do not necessarily reflect the views of the United States Government, DARPA, or NSF.

Kai-Min Chung is supported by Ministry of Science and Technology, Taiwan, under Grant No. MOST 109-2223-E-001-001-MY3.

Giulio Malavolta is supported by the German Federal Ministry of Education and Research BMBF (grant 16K15K042, project 6GEM).

\bibliographystyle{alpha}
\bibliography{cryptobib/abbrev0,cryptobib/crypto_crossref,additionalRef}
\addcontentsline{toc}{section}{References}

\newpage
\appendix
\section*{Supplementary Material}

\section{Additional Preliminaries}
\label{sec:add-prelim}
\subsection{Preliminaries for Lattice}
\label{sec:add-prelim:lattice}
{Throughout the current paper, we denote the Gram-Schmidt ordered orthogonalization of a matrix $\vb{A} \in \mathbb{Z}^{m\times m}$ by $\tilde{\vb{A}}$}.
\subsubsection{Lattices}

We define the notion of a lattice and integer lattice. 

\begin{definition}[Lattice]
Let $\mathbf{B} = [~\mathbf{b}_1~|~\dots~|~\mathbf{b}_m~]$ be a basis of linearly independent vectors $\mathbf{b}_i \in \mathbb{R}^m, \ i \in [m]$. The {\em lattice} generated by $\mathbf{B}$ is defined as $\Lambda = \Set{\mathbf{y} \in \mathbb{R}^m: \exists s_i\in \mathbb{Z}, \mathbf{y} = \sum_1^ms_i\mathbf{b}_i}$. The {\em dual lattice} $\Lambda^*$ of $\Lambda$ is defined as $\Lambda^* = \Set{\mathbf{z} \in \mathbb{R}^m: \forall y \in \Lambda,\langle\mathbf{z,y} \rangle \in \mathbb{Z}}$
\end{definition}

\begin{definition}[Integer Lattice]
For a prime $q$, a modular matrix $\mathbf{A} \in \mathbb{Z}_q^{n\times m}$ and vector $\mathbf{u} \in \mathbb{Z}_q^n$, we define the m-dimensional (full rank) integer lattice $\Lambda_q^\perp(\mathbf{A}) = \Set{\mathbf{e} \in \mathbb{Z}^m:\mathbf{Ae}=0\ (\mod q)}$, and the `shifted' lattice as the coset $\Lambda_q^\mathbf{u}(\mathbf{A}) = \Set{\mathbf{e} \in \mathbb{Z}^m:\mathbf{Ae}=\mathbf{u}\ (\mod q)}$   
\end{definition}

\subsubsection{Lattice Trapdoors, Discrete Gaussians}

The works \cite{STOC:Ajtai96,EC:MicPei12} show how to sample close to uniform matrices $\mathbf{A} \in \mathbb{Z}_q^{n\times m}$ along with a matrix trapdoor $\mathbf{T_A}$ that consists of a basis of low norm vectors for the associated lattice $\Lambda_q^\perp(\mathbf{A})$. We call this sampling procedure $\algo{TrapGen}$. 

\begin{lemma}[Trapdoor Matrices]\label{lemma:trapgen}
There is a PPT algorithm $\algo{TrapGen}$ that given as input integers $n \geq 1$, $q \geq 2$, and (sufficiently large) $m = O(n\log q)$, outputs a matrix $\mathbf{A} \in \mathbb{Z}_q^{n\times m}$ and a trapdoor matrix $\mathbf{T_A} \in \mathbb{Z}_q^{m\times m}$, such that $\mathbf{AT_A} = 0$, the distribution of $\mathbf{A}$ is statistically close to uniform over $\mathbb{Z}_q^{n\times m}$, and $||\widetilde{\vb{T}}_{\vb{A}}|| = O(\sqrt{n\log q})$.  
\end{lemma}

We now define the notion of discrete Gaussian distributions. 

\begin{definition}[Discrete Gaussians]\label{def:DiscGaussian}
Let $m \in \mathbb{Z}_{>0}$, $\Lambda \subset \mathbb{Z}^m$. For any vector $\mathbf{c} \in \mathbb{R}^m$, and positive real $\sigma \in \mathbb{R}_{>0}$, define the Gaussian function $\rho_{\sigma,\mathbf{c}}(\mathbf{x}) = \mathsf{exp}(-\pi||\mathbf{x-c}||^2/\sigma^2)$ over $\mathbb{R}^m$ with center $\mathbf{c}$ and width $\sigma$. Define the discrete Gaussian distribution over $\Lambda$ with center $\mathbf{c}$ and width $\sigma$ as $\mathcal{D}_{\Lambda,\sigma,\mathbf{c}} = \rho_{\sigma,\mathbf{c}}/\rho_{\sigma}(\Lambda)$ where $\rho_{\sigma}(\Lambda) = \sum_{x \in Lambda} \rho_{\sigma,\mathbf{c}}$. For convenience, we use the shorthand $\rho_\sigma$ and $\mathcal{D}_{\Lambda,\sigma}$ for $\rho_{\sigma,\mathbf{0}}$ and $\mathcal{D}_{\Lambda,\sigma,\mathbf{0}}$ respectively. 
\end{definition}

The following lemma is a very useful concentration bound on the norm of discrete guassian samples, depending on the basis they were sampled using. 

\begin{lemma}[Discrete Gaussian Concentration \cite{FOCS:MicReg04}]\label{lemma:discgaussconc}
For any lattice $\Lambda$ of integer dimension $m$ with basis $\mathbf{T}$, $\mathbf{c} \in \mathbb{R}^m$, and Gaussian width parameter $\sigma \geq ||\widetilde{\mathbf{T}}||\cdot\omega\big(\sqrt{\log m}\big)$, we have $$\Pr[\mathbf{x}\gets\mathcal{D}_{\Lambda,\sigma, \vb{c}}: ||\mathbf{x-c}||>\sigma\sqrt{m}] \leq \negl(n)$$ 
\end{lemma}

\subsubsection{The Gadget Matrix}

The gadget matrix $\mathbf{G}$ was defined in \cite{EC:MicPei12}. We use the following two properties of $\mathbf{G}$ in particular: 

\begin{lemma}[{\cite[Theorem 1]{EC:MicPei12}}]\label{lemma:gadgettrapdoor}
Let $q$ be a prime, and $n,m$ be integers with $m=n\log q$. There is a fixed full-rank matrix $\mathbf{G} \in \mathbb{Z}_q^{n\times m}$ such that the lattice $\Lambda_q^\perp(\mathbf{G})$ has a publicly known trapdoor matrix $\mathbf{T_G} \in \mathbb{Z}^{n\times m}$ with $||\widetilde{\vb{T}}_{\vb{G}}|| \leq \sqrt{5}$.
\end{lemma}

\begin{lemma}[{\cite[Lemma 2.1]{EC:BGGHNS14}}]\label{lemma:gadgetinverse}
There is a deterministic algorithm, denoted by $\mathbf{G}^{-1}(\cdot):\mathbb{Z}_q^{n\times m} \rightarrow \mathbb{Z}^{m\times m}$ that takes a matrix $\mathbf{A} \in \mathbb{Z}_q^{n\times m}$ as input, and outputs a `preimage' $\mathbf{G}^{-1}(\mathbf{A})$ of $\mathbf{A}$ such that $\mathbf{G} \cdot \mathbf{G}^{-1}(\mathbf{A})=\mathbf{A}$ $(\mathrm{mod}~q)$ and $||\mathbf{G}^{-1}(\mathbf{A})||\leq m$. 
\end{lemma}

\subsubsection{Hardness Assumptions}
We recall the LWE and SIS problems, and their hardness based on worst case lattice problems.  

For a positive integer dimension $n$ and modulus~$q$, and an error
distribution~$\chi$ over~$\mathbb{Z}$, the LWE distribution and decision
problem are defined as follows.  For an $\mathbf{s} \in \mathbb{Z}^{n}$, the LWE
distribution $A_{\mathbf{s},\chi}$ is sampled by choosing a uniformly
random $\mathbf{a} \gets \mathbb{Z}_q^{n}$ and an error term $e \gets \chi$, and
outputting $(\mathbf{a}, b = \langle\mathbf{s}, \mathbf{a}\rangle + e) \in \mathbb{Z}_q^{n+1}$. 

\begin{definition}
	\label{def:lwe}
	The decision-LWE$_{n,q,\chi}$ problem is to distinguish, with
	non-negligible advantage, between any desired (but polynomially
	bounded) number of independent samples drawn from $A_{\mathbf{s},\chi}$
	for a single $\mathbf{s} \gets \mathbb{Z}_q^{n}$, and the same number of
	\emph{uniformly random} and independent samples over
	$\mathbb{Z}_q^{n+1}$.
\end{definition}

 A standard
instantiation of LWE is to let~$\chi$ be a \emph{discrete Gaussian}
distribution over~$\mathbb{Z}$ with parameter $r = 2\sqrt{n}$.  A sample drawn
from this distribution has magnitude bounded by, say,
$r\sqrt{n} = \Theta(n)$ except with probability at most $2^{-n}$, and
hence this tail of the distribution can be entirely removed. For this
parameterization, it is known that LWE is at least as hard as
\emph{quantumly} approximating certain ``short vector'' problems on
$n$-dimensional lattices, in the worst case, to within
$\widetilde{O}(q\sqrt{n})$
factors~\cite{STOC:Regev05,STOC:PeiRegSte17}.
Classical reductions are also known for different
parameterizations~\cite{STOC:Peikert09,STOC:BLPRS13}.


\begin{definition}\label{def:sis}
The $\mathbf{SIS}_{q,\beta,n,m}$ problem is: given an uniformly random matrix $\mathbf{A} \in \mathbb{Z}_q^{n\times m}$, find a nonzero integral vector $\mathbf{z} \in \mathbb{Z}^m$ such that $\mathbf{Az}=\vb{0}\mod q$, and $||\mathbf{z}|| \leq \beta$. 
\end{definition}

When $q \geq \beta\cdot \widetilde{O}(\sqrt{n})$, solving $\mathbf{SIS}_{q,\beta,n,m}$ is at least as hard as approximating certain worst-case lattice problems (namely, SIVP) to within a $\beta\cdot \widetilde{O}(\sqrt{n})$ factor \cite{FOCS:MicReg04}.

\subsection{Random Sampling Related}
	\label{sec:additional-prelims:sampling-related}
We recall the following generalization of the leftover hash lemma. 

\begin{lemma}[{\cite[Lemma 4]{EC:AgrBonBoy10}}]\label{lemma:lhl}
Suppose that $m > (n+1)\log_2q+\omega(\log n)$ and that $q > 2$ is a prime. Let $\mathbf{R}$ be an $m\times k$ matrix chosen uniformly from $\Set{-1,1}^{m\times k}\mod q$ where $k=k(n)$ is polynomial in $n$. Let $\mathbf{A}$ and $\mathbf{B}$ be matrices chosen uniformly in $\mathbb{Z}_q^{n\times m}$ and $\mathbb{Z}_q^{n\times k}$ respectively. Then for all vectors $\mathbf{w} \in \mathbb{Z}_q^m$, the distribution $(\mathbf{A},\mathbf{AR},\mathbf{R}^\top\mathbf{w})$ is statistically close to the distribution $(\mathbf{A},\mathbf{B},\mathbf{R}^\top\mathbf{w})$.  
\end{lemma}

We will give an argument to show how \Cref{lemma:ext-lhl} follows from this. This goes as follows: assume we start with $(\mathbf{A}',\mathbf{A}'\vb{R},\mathbf{R}^\top\mathbf{w})$. This is statistically close to $(\mathbf{A},\mathbf{A}\vb{R},\mathbf{R}^\top\mathbf{w})$ since $\mathbf{A}$ is sampled uniformly, and $\vb{A}'\sind\vb{A}$. 
By \Cref{lemma:lhl} above, $(\mathbf{A},\mathbf{AR},\mathbf{R}^\top\mathbf{w}) \sind (\mathbf{A},\mathbf{B},\mathbf{R}^\top\mathbf{w})$. The latter is in turn statistically close to $(\mathbf{A}',\mathbf{B},\mathbf{R}^\top\mathbf{w})$. Therefore, we have $(\mathbf{A}',\mathbf{B},\mathbf{R}^\top\mathbf{w}) \sind (\mathbf{A}',\mathbf{A}'\vb{R},\mathbf{R}^\top\mathbf{w})$, concluding the proof for \Cref{lemma:ext-lhl}.

We also recall the following concentration bound on the operator norm for the matrices $\mathbf{R}$. 

\begin{lemma}[{\cite[Lemma 5]{EC:AgrBonBoy10}}]\label{lemma:operatorbd}
Let $\mathbf{R}$ be an uniformly random chosen matrix from $\Set{-1,1}^{m\times m}$, then $\Pr[||\mathbf{R}||_2 > 12\sqrt{2m}] < e^{-m}$. 
\end{lemma}

\subsection{Key-Homomorphic Evaluation Algorithms}
\label{sec:additional-prelims:key-homo-eval}
We recall the matrix key-homomorphic evaluation algorithm from  \cite{C:GenSahWat13,EC:BGGHNS14,ITCS:BraVai14} more fully. This was developed in the context of fully homorphic encryption and attribute-based encryption. This template works generally as follows: given a Boolean {\sf NAND} circuit $C:\Set{0,1}^\ell \rightarrow \Set{0,1}$ with fan-in 2, $\ell$ matrices $\Set{\mathbf{A}_i = \mathbf{AR}_i + x_i\mathbf{G} \in \mathbb{Z}_q^{n\times m}}_{i\in [\ell]}$ which correspond to each input wire of $C$ where $\mathbf{A}\pick \mathbb{Z}_q^{n\times m}$, $\mathbf{R}_i \pick \Set{-1,1}^{m\times m}$, $x_i \in \Set{0,1}$ and $\mathbf{G} \in \mathbb{Z}_q^{n\times m}$ is the gadget matrix, the key-homomorphic evaluation algorithm deterministically computes $\mathbf{A}_C = \mathbf{AR}_C + C(x_1,\dots,x_\ell)\mathbf{G} \in \mathbb{Z}_q^{n\times m}$ where $\mathbf{R}_C \in \Set{-1,1}^{m\times m}$ has low norm and $C(x_1,\dots,x_\ell) \in \Set{0,1}$ is the output bit of $C$ on the arguments $x_1,\dots,x_\ell$. This is done by inductively evaluating each {\sf NAND} gate. For a {\sf NAND} gate $g(u,v;w)$ with input wires $u,v$ and output wire $w$, we have (inductively) matrices $\mathbf{A}_u = \mathbf{AR}_u + x_u\mathbf{G}$, and $\mathbf{A}_v = \mathbf{AR}_v + x_v\mathbf{G}$ where $x_u$ and $x_v$ are the input bits of $u$ and $v$ respectively, and the evaluation algorithm computes 
\begin{equation} 
\begin{split}
 \mathbf{A}_w & = \mathbf{G} - \mathbf{A_u}\cdot\mathbf{G}^{-1}(\mathbf{A}_v) \\
 & = \mathbf{G} - (\mathbf{AR}_u+x_u\mathbf{G})\cdot\mathbf{G}^{-1}(\mathbf{AR}_v + x_v\mathbf{G}) \\
 & = \mathbf{AR}_g + (1-x_ux_v)\mathbf{G}
\end{split}
\end{equation}    
where $1-x_ux_v \coloneqq \algo{NAND}(x_u,x_v)$, and $\mathbf{R}_g = -\mathbf{R}_u\cdot\mathbf{G}^{-1}(\mathbf{A}_v)-x_u\mathbf{R}_v$ has low norm if both $\mathbf{R}_u$ and $\mathbf{R}_v$ have low norm. 

In \cite{ITCS:BraVai14}, Brakerski and Vaikuntanathan observed that the norm of $\mathbf{R}_C$ in the outlined evaluation procedure grows asymmetrically (in the $\mathbf{R}$s corresponding to the input wires). They exploited this observation to design a special evaluation algorithm that evaluates circuits in $\mathsf{NC^1}$ with moderate blowup in the norm of $\mathbf{R}_C$. Specifically, the observation is that any circuit with depth $d$ can be simulated by a length $4^d$ and width 5 branching program by Barrington's theorem, recalled below: 

\begin{theorem}[Barrington's Theorem]\label{theorem:barrington}
Every Boolean {\sf NAND} circuit $C$ that acts on $\ell$ inputs and has depth $d$ can be computed by a width 5 permutation branching program of length $4^d$. Given the description of the circuit $C$, the description of the corresponding branching program can be computed in $\poly(\ell,4^d)$ time. 
\end{theorem}

Such a branching program can then be computed by multiplying $4^d$ many $5\times 5$ permutation matrices. It is shown in \cite{ITCS:BraVai14} that homomorphically evaluating the multiplication of permutation matrices using the above procedure and the asymmetric noise growth feature only increases the noise by a polynomial factor, and thus allows us to use an LWE or SIS modulus that is polynomial in the security parameter. In our constructions, we will use this particular evaluation method just as in \cite{ITCS:BraVai14} and denote it by $\algo{Eval}_{BV}$. 

We will use a claim regarding the noise growth properties of $\algo{Eval}_{BV}$. It can be obtained from Claim 3.4.2 and Lemma 3.6 of \cite{ITCS:BraVai14} and Barrington's Theorem. 

\begin{lemma}\label{lemma:rnorm}
Let $C:\Set{0,1}^\ell \rightarrow \Set{0,1}$ be a {\sf NAND} Boolean circuit. Let $\Set{\mathbf{A}_i = \mathbf{AR}_i + x_i\mathbf{G} \in \mathbb{Z}_q^{n\times m}}_{i\in [\ell]}$ be $\ell$ distinct matrices corresponding to the input wires of $C$, where $\mathbf{A}\pick \mathbb{Z}_q^{n\times m}$, $\mathbf{R}_i \pick \Set{-1,1}^{m\times m}$, $x_i \in \Set{0,1}$ and $\mathbf{G} \in \mathbb{Z}_q^{n\times m}$ is the gadget matrix. There is an efficient deterministic algorithm $\algo{Eval}_{BV}$ that takes as input $C$ and $\Set{\mathbf{A}_i}_{i\in [\ell]}$ and outputs a matrix $\mathbf{A}_C = \mathbf{AR}_C + C(x_1,\dots,x_\ell)\mathbf{G} = \algo{Eval}_{BV}(C,\mathbf{A}_1,\dots,\mathbf{A}_\ell)$ where $\mathbf{R}_C \in \mathbb{Z}^{m\times m}$ and $C(x_1,\dots,x_\ell)$ is the output of $C$ on the arguments $x_1,\dots,x_\ell$. $\algo{Eval}_{BV}$ runs in time $\poly(4^d,\ell,n,\log q)$. 
Let $||\mathbf{R}_{max}||_2 = max\Set{||\mathbf{R}_i||_2}_{i \in [\ell]}$, the norm of $\mathbf{R}_C$ in $\mathbf{A}_C$ output by $\algo{Eval}_{BV}$ can be bounded with overwhelming probability by 
\begin{equation} 
\begin{split}
  ||\mathbf{R}_C||_2 & \leq O(L\cdot ||\mathbf{R}_{max}||_2.m)\\
 & \leq  O(L\cdot 12\sqrt{2m}\cdot m) \\
 & \leq  O(4^dm^{3/2})
\end{split}
\end{equation}     

where $L$ is the length of the width 5 branching program which simulates $C$ and we have used \Cref{lemma:operatorbd} to obtain $||\mathbf{R}_i||_2 \leq 12\sqrt{2m}$ for each $i$ with overwhelming probability. 
In particular, if $C$ is in $\mathsf{NC^1}$ and has depth $d=c\log l$ for a constant $c$, then $L=4^d=\ell^{2c}$ and $\leq  O(\ell^{2c}\cdot m^{3/2})$
\end{lemma}

\section{One-More Unforeagibility vs PQ-EUF for Ring Signatures}
\label{sec:one-more:PQ-EUF:ring-sig}

The ring-signature analog of the one-more unforgeability by Boneh and Zhandry \cite{C:BonZha13}, {\em when restricted to the classical setting}, seems to be weaker than the standard unforgeability in \Cref{def:classical:ring-signature}.\footnote{This is in contrast to the case of ordinary signatures, where one-more unforgeability is equivalent to the standard existential unforgeability \cite{C:BonZha13}.} That is, in the classical setting, any $\RS$ satisfying the unforgeability in \Cref{def:classical:ring-signature} is also one-more unforgeable; but the reverse direction is unclear. We provide discussion in the following.

To argue that one-more unforgeability is no weaker than \Cref{def:classical:ring-signature}, one needs to show how to convert a forger $\Adv_\textsc{euf}$ winning in \Cref{expr:classical:unforgeability} to another forger $\Adv_\textsc{om}$ winning in the (classical version of) ``one-more forgery'' game. Conceivably, $\Adv_\textsc{om}$ will run $\Adv_\textsc{euf}$ internally; thus, $\Adv_\textsc{om}$ will make no less {\sf sign} queries than $\Adv_\textsc{euf}$. Recall that $\Adv_\textsc{om}$ needs to forge one more signature than the total number of its queries.
 Also, crucially, all the ring signatures presented by $\Adv_\textsc{om}$ at the end must have {\em no} corrupted members in the accompanying ring. Now ideally one might imagine that we can simply use the queries made by $\Adv_\textsc{om}$ (which are really queries by $\Adv_\textsc{euf}$) to meet the ``one-more'' challenge; however, this is thwarted immediately due to the fact that $\Adv_\textsc{euf}$ has absolutely no obligation to make queries meeting this requirement, so even if the final forgery produced by $\Adv_\textsc{euf}$ is valid, our attempted reduction does not have any means to provide $\Adv_\textsc{om}$ with all the signatures it needs to win the ``one-more'' challenge (since not all of the queries can be reused). Indeed, it is not hard to find attacks that use this definitional gap to violate standard unforgeability, while being ruled out as a valid attack against one-more ring unforgeability. Contrast this with a comparison in the other direction: an adversary $\Adv_\textsc{om}$ for the one-more unforgeability experiment is easily converted into a standard $\Adv_\textsc{euf}$ adversary since not all of the signatures output by $\Adv_\textsc{om}$ at the end can be previous queries (by the pigeonhole principle); $\Adv_\textsc{euf}$ simple outputs the one that is not. 

We remark however that this definitional gap between standard ring signature unforgeability and the ``one-more'' version may not be inherent; rather, we just do not know how to meet this gap. Our arguments here should not be interpreted as a proof showing that the former notion is strictly stronger than the latter. We leave it as an open question to either demonstrate a separation, or prove that the two are actually equivalent.

\end{document}